\documentclass[preprintnumbers,article,amsmath,amssymb,floatfix,10pt,prd,twocolumn,
superscriptaddress,nofootinbib]{revtex4-2}
\usepackage{bm}
\usepackage{amsfonts}
\usepackage{latexsym}
\usepackage[latin1]{inputenc}
\usepackage{graphicx}
\usepackage{amsmath}
\usepackage{palatino}
\usepackage{mathpazo}
\usepackage{textcomp}
\usepackage{float}
\usepackage{booktabs}
\usepackage{dcolumn}
\usepackage{ragged2e}
\usepackage{hyperref}
\hypersetup{colorlinks,citecolor=blue}
\hypersetup{colorlinks=true,linkcolor=red,filecolor=magenta,    urlcolor=blue}
\usepackage{amsmath}
\usepackage{xcolor}
\usepackage{orcidlink}
\usepackage{epsfig}
\usepackage{subfigure}
\usepackage{commath}

\def\jnl@style{\it}
\def\aaref@jnl#1{{\jnl@style#1}}

\def\aaref@jnl#1{{\jnl@style#1}}

\def\aj{\aaref@jnl{AJ}}                   % Astronomical Journal
\def\apj{\aaref@jnl{ApJ}}                 % Astrophysical Journal
\def\apjl{\aaref@jnl{ApJ}}                % Astrophysical Journal, Letters
\def\apjs{\aaref@jnl{ApJS}}               % Astrophysical Journal, Supplement
\def\apss{\aaref@jnl{Ap\&SS}}             % Astrophysics and Space Science
\def\aap{\aaref@jnl{A\&A}}                % Astronomy and Astrophysics
\def\aapr{\aaref@jnl{A\&A~Rev.}}          % Astronomy and Astrophysics Reviews
\def\aaps{\aaref@jnl{A\&AS}}              % Astronomy and Astrophysics, Supplement
\def\mnras{\aaref@jnl{Mon.~Not.~Roy.~Astron.~Soc.}}             % Monthly Notices of the RAS
\def\prd{\aaref@jnl{Phys.~Rev.~D}}        % Physical Review D
\def\prc{\aaref@jnl{Phys.~Rev.~C}}  % Physical Review C
\def\prl{\aaref@jnl{Phys.~Rev.~Lett.}}    % Physical Review Letters
\def\qjras{\aaref@jnl{QJRAS}}             % Quarterly Journal of the RAS
\def\skytel{\aaref@jnl{S\&T}}             % Sky and Telescope
\def\ssr{\aaref@jnl{Space~Sci.~Rev.}}     % Space Science Reviews
\def\zap{\aaref@jnl{ZAp}}                 % Zeitschrift fuer Astrophysik
\def\nat{\aaref@jnl{Nature}}              % Nature
\def\aplett{\aaref@jnl{Astrophys.~Lett.}} % Astrophysics Letters
\def\apspr{\aaref@jnl{Astrophys.~Space~Phys.~Res.}} % Astrophysics Space Physics Research
\def\physrep{\aaref@jnl{Phys.~Rep.}}      % Physics Reports
\def\physscr{\aaref@jnl{Phys.~Scr}}       % Physica Scripta
\def\commat{\aaref@jnl{Comm.~Math.~Phys.}}              % Communications in Mathematical Physics
\def\science{\aaref@jnl{Science}}               % Science
\def\cqg{\aaref@jnl{Classical Quant.~Grav.}}            % Classical and Quantum Gravity
\def\jpcs{\aaref@jnl{JPCS}}                                     % Journal of Physics Conference Series
\def\ijmpd{\aaref@jnl{Int.~J.~Mod.~Phys.~D}}                    % International Journal of Modern Physics D
\def\grg{\aaref@jnl{Gen.~Relat.~Gravit.}}               % General Relativity and Gravitation
\def\rpp{\aaref@jnl{Rep.~Prog.~Phys.}}          % Reports on Progress in Physics
\def\npa{\aaref@jnl{Nucl.~Phys.~A}}        % Nuclear Physics A
\def\lrr{\aaref@jnl{Living Rev.~Rel.}}                   % Living reviews in relativity
\def\jcap{\aaref@jnl{J.~Cosmology Astropart.~Phys.}}    % Journal of cosmology and astroparticle physics
\def\rmp{\aaref@jnl{Rev.~Mod.~Phys.}}   %Reviews of modern physics
\def\epjc{\aaref@jnl{Eur.~Phys.~J.~C}} 
\def\plb{\aaref@jnl{~Phy.~Lett.~B}} 
\def\mpla{\aaref@jnl{Mod.~Phy.~Lett.~A}} 
\def\arxiv{\aaref@jnl{arxiv.org}}

\allowdisplaybreaks[1]
\begin{document}
%\color{red}
\color{black}       %% For one column
\title{Casimir Wormholes in Modified Symmetric Teleparallel Gravity}

\author{Zinnat Hassan\orcidlink{0000-0002-6608-2075}}
\email{zinnathassan980@gmail.com}
\affiliation{Department of Mathematics, Birla Institute of Technology and
Science-Pilani,\\ Hyderabad Campus, Hyderabad-500078, India.}

\author{Sayantan Ghosh\orcidlink{0000-0002-3875-0849}}
\email{sayantanghosh.000@gmail.com}
\affiliation{Department of Mathematics, Birla Institute of Technology and
Science-Pilani,\\ Hyderabad Campus, Hyderabad-500078, India.}

\author{P.K. Sahoo\orcidlink{0000-0003-2130-8832}}
\email{pksahoo@hyderabad.bits-pilani.ac.in}
\affiliation{Department of Mathematics, Birla Institute of Technology and
Science-Pilani,\\ Hyderabad Campus, Hyderabad-500078, India.}

\author{Kazuharu Bamba\orcidlink{0000-0001-9720-8817}}
\email{bamba@sss.fukushima-u.ac.jp}
\affiliation{Faculty of Symbiotic Systems Science, 
Fukushima University, Fukushima 960-1296, Japan}

%
%%%%%%%%%%%%%%%%%%%%%%%%%%%%%%%%%%%%%  DATE  %%%%%%%%%%%%%%%%%%%%%%%%%%%%%%%%%%%%
\date{\today}
\begin{abstract}
In recent years there has been a growing interest in the field of Casimir wormhole. In classical general relativity (GR), it is known that the null energy condition (NEC) has to be violated to have a wormhole to be stable. The Casimir effect is an experimentally verified effect that is caused due to the vacuum field fluctuations in quantum field theory. Since the Casimir effect provides the negative energy density, thus this act as an ideal candidate for the exotic matter needed for the stability of the wormhole. In this paper, we study the Casimir effect on the wormhole geometry in modified symmetric teleparallel gravity or $f(Q)$ gravity, where the non-metricity scalar $Q$ drives the gravitation interaction. We consider three systems of the Casimir effect such as (i) two parallel plates, (ii) two parallel cylindrical plates, and (iii) two-sphere separated by a large distance to make it more experimentally feasible. Further, we studied the obtained wormhole solutions for each case with energy conditions at the wormhole throat with radius $r_0$ and found that some arbitrary quantity violates the classical energy conditions at the wormhole throat. Furthermore, the behavior of the equation of state (EoS) is also analyzed for each case. Finally, we investigate the stability of the obtained Casimir effect wormhole solutions with the generalized Tolman-Oppenheimer-Volkoff (TOV) equation.
\end{abstract}

\maketitle

%\date{\today}

\section{Introduction}\label{sec1}
In 1915, Albert Einstein proposed the General Theory of Relativity (GR), which aims to express the gravitational interaction as a geometric effect of space-time in light of a matter source \cite{Padmanabhan,Misner/1973}. Over time, it has been demonstrated to be very emphasized, providing predictability of different phenomena \cite{Dyson,Suzuki,Abbott}. Karl Schwarzschild, in 1916 found the first non-trivial analytic solution of Einstein GR in \cite{Schwarzschild}. Also, solution of Schwarzschild displays a removable singularity at a regular point, called the event horizon. This concept provided the idea of the black hole, and recently its existence has been proved in \cite{Akiyama/2019,Akiyama/2021}. Interestingly, in the same year, 1916, Wheeler \cite{Flamm} discovered another possible solution nowadays known as \textit{white hole}.\\
Einstein and Rosen, in 1935, first adopted the concept of GR and made a hypothetical bridge called the Einstein-Rosen bridge or wormholes through space-time \cite{Rosen}. Basically, wormholes are used to connect two asymptotically flat different regions of space-time by a throat \cite{Visser}. Nevertheless, the wormhole's existence requires to be observed experimentally. Morris and Thorne first introduced the idea of a traversable wormhole in 1988 \cite{Thorne/1988}. Two geometric conditions need to be imposed to maintain the wormhole's traversability: the flare-out condition \cite{Hochberg} (it is affiliated to minimization of the wormhole throat) and the finite redshift function (to avoid the event horizon). To support the construction of the wormhole, a hypothetical type of matter requires so-called exotic matter, which obeys the flare-out condition and disrespects the weak energy condition (WEC) \cite{Yurtsever}. However, the author of Ref. \cite{Kim} demonstrated using the embedding procedure that negative energy density fulfills the flare-out condition for the Morris-Thorne (MT) type wormhole, although it is not mandatory to violate the WEC \cite{Simeone}. Understanding such solutions has evolved into the topic of enthusiastic research in recent years \cite{Gao}, including the finding of humanly traversable wormholes in Ref. \cite{Milekhin}.\\
It is known that GR describes the universe well enough; however, some recent observations demand modifying the classical relativity into some alternative that could explain well, such as the late-time acceleration expansion of the universe \cite{Perlmutter,Komatsu,Riess,Suzuki/2012}. Thus, different modified theories have been introduced to overpass these issues (for examples $f(R)$ gravity \cite{Buchdahl/1970,Starobinsky/1980}, $f(R,T)$ gravity \cite{Lobo/2011}, etc). Numerous articles are available on wormhole geometry in the field of modified theories of gravity \cite{Lobo/2009,Bronnikov/2010,Kar/2003,Banerjee,Bambi/2015,Ovgun,Sahoo1,Karar1}. Recently, some interesting works on Kerr-like wormholes and the shadows of wormholes have been investigated in \cite{Shaikh,Nedkova,Tsukamoto,Amir}, also in $f(R)$ \cite{Bahamonde2} and $f(T)$ gravity \cite{Bahamonde}.
\\
In recent years there has been some development in the Casimir wormhole. It is well known that for the wormhole to be stable, we need to violate the Null Energy Condition (NEC), which can only be done by so-called exotic matter. One practical example of such matters can be found in Casimir effect \cite{casimir} in quantum field theory, where we can get tap into infinite vacuum energy by just putting two parallel conducting plates. A similar calculation was also done by Lifshitz etc. \cite{lifshitz} via a little different route but yielded the same result. In 1997 it was shown experimentally by Lamoreaux \cite{experiment} confirming the earlier assertion about the effect. Later the experiment was improved, and Bressi \cite{bressi} obtained an even better result. \\
Even though results for the Casimir effect for parallel plates are well known, it becomes very difficult to calculate the Casimir force when the shape of the conductors is different are rather hard to compute. Fortunately, there are some cases where the exact results are known, like two concentric cylinders or two concentric spheres, but to handle the general cases, one needs to approach some kind of approximate method to get the result. Two popular approximation methods are PFA and the other being the multiple scattering approach. For this paper, we have taken three energy densities to calculate the various quantities of wormholes that is the parallel plate, the concentric cylinder, and two spheres separated by a distance.\\
Later the calculation for Casimir effects was extended two add more nontrivial geometry via the method of Proximity Force Approximation(PFA) first done by \cite{PFA}. The idea of PFA was that even if the plates are not completely parallel to each other, we can approximately use this formula to calculate the energy
$$E_{PFA}=\int_{\Sigma}d\sigma E_{pp}(Z)$$
where $E_{pp}=\frac{\pi^2}{720}\frac{S}{a^3}$ and $\Sigma$ is one of the surfaces (under PFA approximation, it does not matter which one) and $d\sigma$ denotes the area of the surface.\\
This article aims to study the Casimir effect in wormhole space-time geometry. Here, we investigate three different cases of the Casimir effect such as (i) parallel plates, (ii) two concentric cylinders (commonly known as the "rack and pinion" model), and (iii) two spheres, and discuss the behavior of shape functions and energy conditions in modified symmetric teleparallel gravity or $f(Q)$ gravity.\\
%In \cite{Garattini}, Garattini proposed the Generalized Uncertainty Principle (GUP) Casimir wormhole. But, in this article, we have not investigated the effect of GUP on Casimir wormholes.
The $f(Q)$ gravity (where $Q$ is the non-metricity scalar) is a recently proposed non-metricity-based modified gravity proposed by Jimenez et al. \cite{Jimenez}. Although, it is recently proposed gravity, many interesting works have already been done, such as, cosmography \cite{Mandal/2020}, energy conditions \cite{Mandal/2020a},  covariant formulation \cite{Zhao}, signature of $f(Q)$ gravity \cite{Frusciante} etc. The coupling matter in modified $Q$ gravity has been analyzed by Harko et al. in \cite{Harko/2018a}. Also, the authors of Ref. \cite{Khyllep1} discussed The growth index of matter perturbations in this modified gravity. Another interesting work by Anagnostopoulos et al. \cite{Anagnostopoulos}, where they provide evidence that non-metricity $f(Q)$ gravity could challenge the $\Lambda$CDM model. Moreover, in the field of astrophysical objects, some interesting literature are there in $f(Q)$ gravity such as in black holes \cite{Fell} wormhole geometry \cite{Hassan1,Mustafa,Sharma1,Banerjee1}, compact star \cite{Mustafa2} and so on. One may check the Refs. \cite{Lazkoz,Lin,Solanki1,Wang2}, where the authors have done some interesting works on cosmological and astrophysical aspects as well.\\
These inspired us to study the Casimir effect in wormholes geometry in $f(Q)$ gravity. This paper is organized as follows: In Sec. \ref{sec2}, we have introduced the basic formalism of $f(Q)$ gravity and also constructed the field equations of MT wormhole metric for the corresponding gravity. The Casimir effect has been discussed in the Sec. \ref{sec3}. Further, in the same section, we consider three different cases of the Casimir effect and compare them with the energy density of wormhole metric and study the behavior with energy conditions. Moreover, in Sec. \ref{sec4}, we checked our obtained wormhole solutions with the TOV equation. We concluded our results in Sec. \ref{sec5}.

\section{wormholes in modified theory of gravity}
\label{sec2}
This section will provide the basic formalism of $f(Q)$ gravity and develop the field equations for this modified gravity. In the following sections, we will study the Casimir effect of wormhole solutions with energy conditions.
\subsection{The $f(Q)$ gravity}
In this subsection, we are going to briefly present some generalities about the $f(Q)$ gravity. The so-called symmetric teleparallel gravity or $f(Q)$ gravity was introduced by Jimenez et al. \cite{Jimenez}. The action for this gravity can be define by
\begin{equation}\label{6a}
\mathcal{S}=\int\frac{1}{2}\,f(Q)\sqrt{-g}\,d^4x+\int \mathcal{L}_m\,\sqrt{-g}\,d^4x\, ,
\end{equation}
where, $f(Q)$ denoted as the arbitrary function of $Q$ and $\mathcal{L}_m$ is Lagrangian density of matter. $g$ is the determinant of the metric tensor $g_{\mu\nu}$.\\
The nonmetricity tensor is define by\\
\begin{equation}\label{6b}
Q_{\lambda\mu\nu}=\bigtriangledown_{\lambda} g_{\mu\nu}=\partial_\lambda g_{\mu\nu}-\Gamma^\beta_{\,\,\,\lambda \mu}g_{\beta \nu}-\Gamma^\beta_{\,\,\,\lambda \nu}g_{\mu \beta},
\end{equation}
where, $\Gamma^\beta_{\,\,\,\mu\nu}$ is the metric affine connection.\\
Also, the superpotential or the non-metricity conjugate can be defined in terms of the nonmetricity tensor as
\begin{multline}\label{6c}
P^\alpha\;_{\mu\nu}=\frac{1}{4}\left[-Q^\alpha\;_{\mu\nu}+2Q_{(\mu}\;^\alpha\;_{\nu)}+Q^\alpha g_{\mu\nu}-\tilde{Q}^\alpha g_{\mu\nu}\right.\\\left.
-\delta^\alpha_{(\mu}Q_{\nu)}\right],
\end{multline}
where
\begin{equation}
\label{6d}
Q_{\alpha}=Q_{\alpha}\;^{\mu}\;_{\mu},\; \tilde{Q}_\alpha=Q^\mu\;_{\alpha\mu}.
\end{equation}
are two independence traces.\\
The nonmetricity scalar represented as \cite{Jimenez}
\begin{equation}
\label{6e}
Q=-Q_{\alpha\mu\nu}\,P^{\alpha\mu\nu}.
\end{equation}
In order to find the field equations for this theory of gravity, we very the action \eqref{6a} with respect to the metric tensor $g_{\mu\nu}$, resulting in
\begin{multline}\label{6f}
\frac{2}{\sqrt{-g}}\bigtriangledown_\gamma\left(\sqrt{-g}\,f_Q\,P^\gamma\;_{\mu\nu}\right)+\frac{1}{2}g_{\mu\nu}f \\
+f_Q\left(P_{\mu\gamma i}\,Q_\nu\;^{\gamma i}-2\,Q_{\gamma i \mu}\,P^{\gamma i}\;_\nu\right)=-T_{\mu\nu},
\end{multline}
where $f_Q=\frac{df}{dQ}$, and $T_{\mu\,\nu}$ is the standard energy-momentum tensor, whose form is
\begin{equation}\label{6g}
T_{\mu\nu}=-\frac{2}{\sqrt{-g}}\frac{\delta\left(\sqrt{-g}\,\mathcal{L}_m\right)}{\delta g^{\mu\nu}}.
\end{equation}
Moreover, by varying the action over the connection, we are able to derive the extra constraint
\begin{equation}\label{6h}
\bigtriangledown_\mu \bigtriangledown_\nu \left(\sqrt{-g}\,f_Q\,P^\gamma\;_{\mu\nu}\right)=0.
\end{equation}
Also, one may study this theory with a special coordinate
choice, the so-called coincident gauge. In that case, the connection will vanish. Thus, the non-metricity \eqref{6b} reduces to
\begin{equation}\label{2222}
Q_{\lambda\mu\nu}=\partial_\lambda g_{\mu\nu}.
\end{equation}
 This simplifies the calculation since only the metric is the fundamental variable. However, in this case, the action no longer remains diffeomorphism invariant, except for standard GR \cite{Koivisto}.
\subsection{Wormholes in $f(Q)$ gravity}
Let us consider the static and spherically symmetric Morris-Thorne wormhole metric  \cite{Visser,Thorne/1988}, define by
\begin{equation}\label{3a}
ds^2=-e^{2\phi(r)}dt^2+\left(1-\frac{b(r)}{r}\right)^{-1}dr^2+r^2d\theta^2+r^2\text{sin}^2\theta d\Phi^2,
\end{equation}
where, $\phi(r)$ and $b(r)$ denoted as the redshift function and the shape function, respectively \cite{2a} and both are functions of radial coordinate $r$. The radial coordinate $r$ decreases from infinity to a minimum radius $r_0$ ($r_0$ is the throat radius), where $b(r_0)=r_0$, then it increases to infinity. An important criteria of a traversable wormholes is the flaring out condition which is given by $(b-b'r)/b^2>0$ \cite{Thorne/1988} and at the throat of the wormhole this condition will reduce to $b^{\,\prime}(r_0)<1$. Also, another condition, for $r>r_0$, $1-\frac{b(r)}{r}>0$ need to be satisfied. In classical GR, these conditions may confirms the presence of exotic matter which requires the NEC violation. Also, to avoid the presence of event horizon, one must demand the redshift function should be finite everywhere.\\
In this work, we study with the energy-momentum tensor for an anisotropic fluid which is define by
\begin{equation}\label{3b}
T_{\mu\,\nu}=\left(\rho+P_t\right)u_{\mu}\,u_{\nu}+P_t\,\delta_{\mu\,\nu}+\left(P_r-P_t\right)v_{\mu}\,v_{\nu},
\end{equation}
where, $u_{\mu}$ and $v_{\mu}$ represents the the four-velocity vector and the unitary space-like vector, respectively and they satisfy the conditions $u_{\mu}u^{\nu}=-v_{\mu}v^{\nu}=-1$. $\rho$ denoted as the energy density. $P_r$ and $P_t$ define as the radial and tangential pressure, respectively and both are function of radial coordinate $r$ only.\\
Using Eq. \eqref{6e}, we can determine the nonmetricity scalar for the metric \eqref{3a} is given by
\begin{equation}\label{6i}
Q=-\frac{b}{r^2}\left(\frac{r b'-b}{r (r-b)}+2 \phi^{'}\right)
\end{equation}
One can note that the above non-metricity scalar will be different if we assume the coincident gauge connection. The authors of \cite{Hassan1,Mustafa,Wang2} considered this special connection in their study on this $f(Q)$ gravity.\\
Now, the field equations for the wormhole metric \eqref{3a} under anisotropic matter source \eqref{3b} in modified symmetric teleparallel gravity can be obtained as
\begin{multline}
\label{11}
\rho =\frac{(r-b)}{2 r^3}\left[\,r\,f_{QQ}Q^{'}\frac{b}{r-b}\right.\\\left.
+f_Q \left(\frac{(2 r-b) \left(r b^{'}-b\right)}{(r-b)^2}+\frac{b \left(2 r \phi^{'}+2\right)}{r-b}\right)+\frac{f r^3}{r-b}\right],
\end{multline}
\begin{multline}
\label{12}
P_r=-\frac{\left(r-b\right)}{2 r^3}\left[\,r\,f_{QQ}Q^{'}\frac{b}{r-b}\right.\\\left.
+f_Q \left(\frac{b}{r-b}\left(\frac{r b{'}-b}{r-b}+2 r \phi^{'}+2\right)-4 r \phi^{'}\right)+\frac{f r^3}{r-b}\right],
\end{multline}
\begin{multline}
\label{13}
P_t=-\frac{\left(r-b\right)}{4 r^2}\left[-4\,r\,\phi^{'} f_{QQ} Q^{'}\right.\\\left.
+ f_Q \left(\frac{\left(r b^{'}-b\right)}{r (r-b)}\left(\frac{2 r}{r-b}+2 r \phi^{'}\right)+\frac{4 (2 b-r) \phi^{'}}{r-b}\right.\right.\\\left.\left.
-r \left(2 \phi^{'}\right)^2-4 r \phi^{''}\right)+\frac{2 f r^2}{r-b}\right],
\end{multline}
where ${'}$ represents $\frac{d}{dr}$. In this study we are working with the simplest linear form of $f(Q)$ gravity such as $f(Q)=\alpha Q+\beta$ \cite{Solanki}. Therefore, we get the revised field equations as follows\\
\begin{equation}
\label{14}
\rho=\frac{\alpha  b'}{r^2}+\frac{\beta }{2},
\end{equation}
\begin{equation}
\label{15}
P_r=\frac{1}{r^3}\left[2 \alpha  r \left(r-b\right) \phi '-\alpha  b\right]-\frac{\beta }{2},
\end{equation}
\begin{multline}
\label{16}
P_t=\frac{1}{2 r^3}\left[\alpha  \left(r \phi '+1\right) \left(-r b'+2 r (r-b) \phi '+b\right)\right]\\
+\frac{\alpha  (r-b) \phi ''}{r}-\frac{\beta }{2}.
\end{multline}\\
Now, with the above expressions (\ref{14}-\ref{16}), we try to find the wormhole solutions under different Casimir sources and study the behavior of those solutions with various tools.\\
Let us devote a couple of words to classical energy conditions developed from the Raychaudhuri equations. Since we are studying with anisotropic fluid matter distribution, the energy condition retrieved from standard GR are\\
$\bullet$ Null energy condition (NEC) if $\rho+P_r\geq0$, \quad  $\rho+P_t\geq0$.\\
$\bullet$ Weak energy conditions (WEC) if $\rho\geq0$, \quad $\rho+P_r\geq0$, \quad  $\rho+P_t\geq0$.\\
$\bullet$ Dominant energy conditions (DEC) if $\rho-\mid P_r\mid \geq0$, \quad  $\rho+\mid P_t\mid \geq0$.\\
$\bullet$ Strong energy conditions (SEC) if $\rho+P_r+2P_t\geq0$.\\
In the context of GR, wormhole disrespect these energy conditions (especially NEC). Taking this in mind, we are going to analyze these energy conditions by defining\\
$NEC1=\rho+P_r$, \quad $NEC2=\rho+P_t$, \quad $DEC1=\rho-\mid P_r \mid$, \quad $DEC2=\rho-\mid P_t \mid$, \quad $SEC=\rho+P_r+2P_t$\\
for our study.\\
 Also, in this work, we consider the constant redshift function i.e., $\phi'(r)=0$, to avoid the event horizon issue.
\section{The Casimir effect}
\label{sec3}
The Casimir effect is a direct manifestation of the fact that in quantum field theory, the vacuum is not really a "vacuum," but it is filled with infinite energy coming from the ground states of the simple harmonic oscillator. As typical scattering processes measure the difference between ground state energy, so those are not really helpful in probing the infinite vacuum energy. It was first \cite{casimir} and independently \cite{lifshitz} who were first able to show that if you put two conducting parallel uncharged plates at a very close distance(closer than the characteristic length scale of the dimension of the plate), then this plates can indeed work as a potential wall and if one sum over the modes one can certainly find the non-trivial energy density. From density, one can find the effective pressure or effective force and then see with experiment whether they match or not as done in \cite{experiment}.\\
There are mainly two similar ways of counting the normal modes for the Casimir effect; as the summation of normal modes leads to summing a divergent series, one has to use either a mathematical cut-off via zeta function regularization \cite{paddy} or via introducing a physical cut off \cite{zee}. Both of them give the same results. That is,\\
 \begin{equation}\label{4a1}
   E(a)=-\frac{\pi^2}{720}\frac{S}{a^3},
 \end{equation}
 where $S$ is the area of the plates, and $a$ is the separation between the plates. This calculation can be performed by adding all the normal modes of the field.\\
 The reason why the Casimir effect is so important in the context of the wormhole is that it violates the NEC, which is necessary for the wormhole to be stable \cite{Thorne/1988}.\\
 \subsection{Case-I: parallel plates}\label{subsec1}
\begin{figure}[h]
%\centering
\includegraphics[scale=0.6]{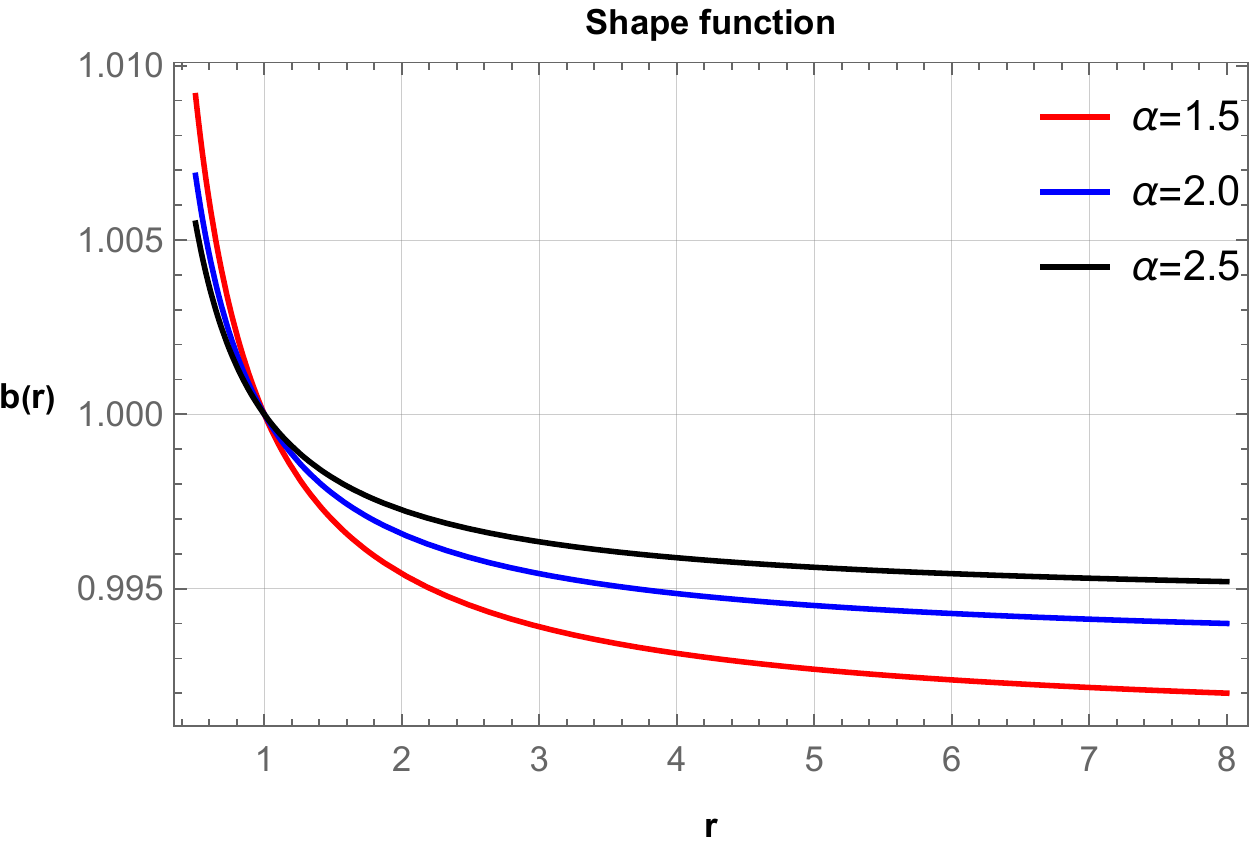}
\includegraphics[scale=0.6]{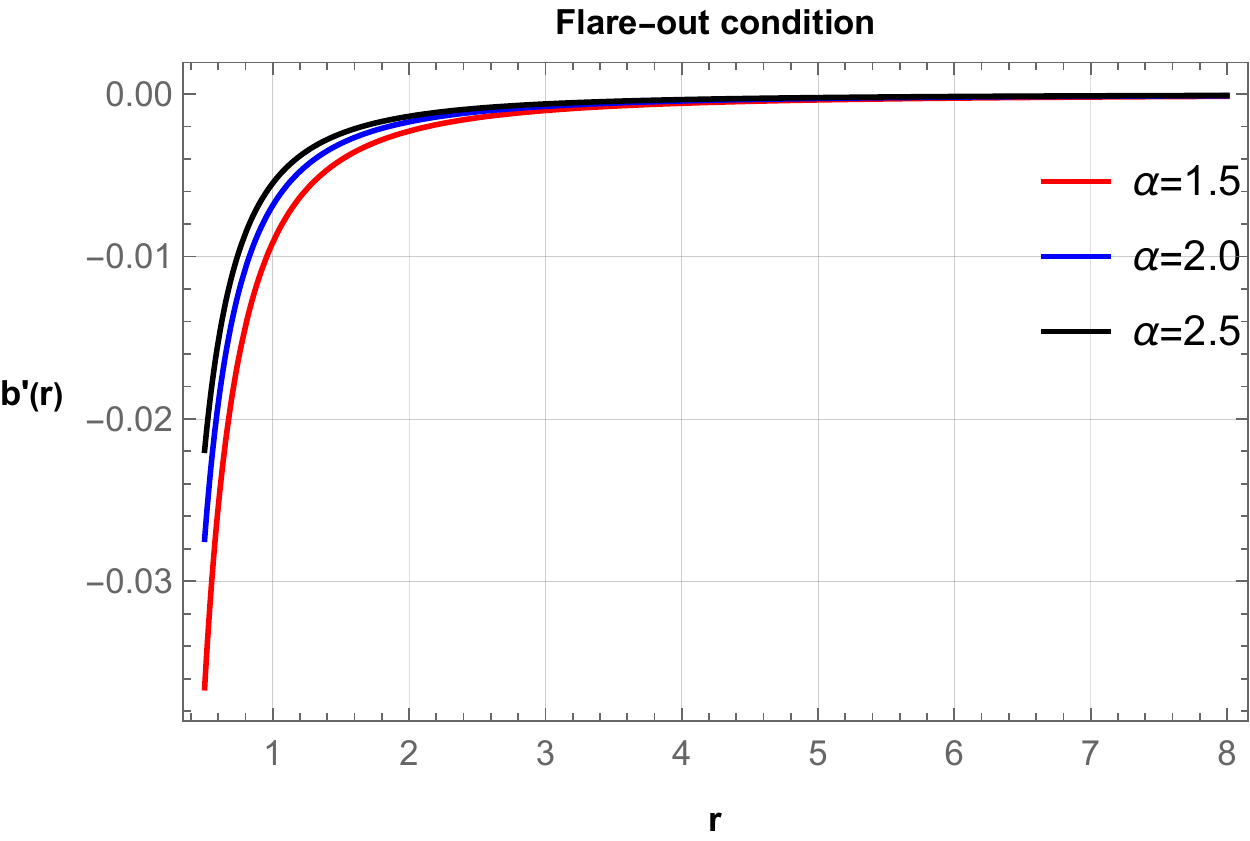}
\caption{Shape function and flare-out condition for parallel plates with $r_0=1$.}
\label{fig6a}
\end{figure}

As we have discussed earlier, this is done in various cases of the wormhole. For this case, we have taken motivation from \cite{Santos}, and in this paper, we are extending the work in the $f(Q)$ model.\\
The Casimir force for the two parallel plates is given by the equation \eqref{4a1}. The energy density $\rho$ is given by the fact that if we bring the two plates closer the energy density is given by the fact that,
\begin{equation}\label{4a2}
   \rho(a)=-\frac{\pi^2}{720}\frac{1}{a^4} 
\end{equation}
We can also get the pressure by the thermodynamic relation
$$p(a)=-\frac{1}{S}\frac{dE(a)}{da}=-\frac{\pi^2}{240}\frac{1}{a^4}$$
So the equation of state is given by
$$p=3\rho$$
Comparing with the general EoS $P=\omega\rho$, we can deduce that the $\omega=3$. In this case, we replace distance $a$ by radial coordinate $r$. Such a choice favours wormhole throat of Planckian scale \cite{Garattini}.\\
Comparing Equations \eqref{14} and \eqref{4a2} and integrating, we get the following form of shape function
\begin{equation}\label{4a3}
    b(r)=-\frac{1}{720 \alpha }\left(120 \beta  r^3-\frac{\pi ^2}{r}\right)+c_1,
\end{equation}
where $c_1$ is the integrating constant and it can be obtained by imposing the throat condition $b(r)=r_0$ at $r=r_0$
\begin{equation}\label{4a4}
    c_1=r_0+\frac{1}{720 \alpha }\left(120 \beta  r_0^3-\frac{\pi ^2}{r_0}\right).
\end{equation}
Inserting Eq. \eqref{4a4} into Eq. \eqref{4a3}, we obtain the shape function as follows:
%\begin{equation}
%\label{4a5}
%b(r)=\frac{120 \beta  \left(r_0^3-r^3\right)+\pi ^2 %\left(\frac{1}{r}-\frac{1}{r_0}\right)}{720 \alpha }+r_0.
%\end{equation}
\begin{equation}
\label{4a5}
b(r)=r_0+\frac{\pi ^2}{720 \alpha}\left(\frac{1}{r}-\frac{1}{r_0}\right)+\frac{\beta}{6 \alpha}\left(r_0^3-r^3\right).
\end{equation}
One may observe that the above equation is not asymptotically flat, i.e., for $r\rightarrow \infty$, $\frac{b(r)}{r}\nrightarrow 0$. It happens because of the third term of the above equation. But, if we consider $\beta \rightarrow 0$, it will satisfy the flatness condition. Thus, to satisfy the asymptotic flatness condition, we are bound to consider vanishing $\beta$ for this case. Thus the above expression reduces to
\begin{equation}
b(r)=r_0+\frac{\pi ^2}{720 \alpha}\left(\frac{1}{r}-\frac{1}{r_0}\right).
\end{equation}
\\
Fig. \ref{fig6a} shows the behavior of shape functions for parallel plates for different possible parameters. In this particular subsection, we have considered the throat radius $r_0=1$. It is observed that the shape function $b(r)$ shows positively increasing behavior for different positively increasing values of $\alpha$ and also satisfies the flaring out conditions, i.e., $b^{'}(r)<1$ at wormhole throat under asymptotically flatness background. For negative $\alpha$, these conditions will not validate.\\
The radial and tangential pressure for the parallel plates in the symmetric teleparallel gravity become
%\begin{equation}
%\label{4a6}
%P_r=-\frac{2 \beta  r^3+\beta  r_0^3+6 \alpha  r_0}{6 r^3}-\frac{\pi ^2 \left(r^2+r_0^2\right)}{720 r^5 r_0^2},
%\end{equation}
\begin{equation}
\label{4a6}
P_r=\frac{\pi ^2 (r-r_0)}{720 r^4 r_0}-\frac{\alpha  r_0}{r^3},
\end{equation}
\begin{equation}
\label{4a7}
P_t=\frac{\alpha  r_0}{2 r^3}-\frac{\pi ^2 (r-2 r_0)}{1440 r^4 r_0}.
\end{equation}
Let us now define a radial EoS parameter $\omega_r(r)=\frac{p_r}{\rho}$ which can be obtained as
\begin{equation}
\label{4a8}
\omega_r(r)=\frac{720 \alpha  r r_0}{\pi ^2}-\frac{r}{r_0}+1,
\end{equation}
Similarly for a scenario with the EoS parameter $\omega_t(r)=\frac{p_t}{\rho}$ which can be obtained as
\begin{equation}
\label{4a9}
\omega_t(r)=-\frac{360 \alpha  r r_0}{\pi ^2}+\frac{r}{2 r_0}-1.
\end{equation}
\begin{figure}[h]
    \centering
    \includegraphics[scale=0.6]{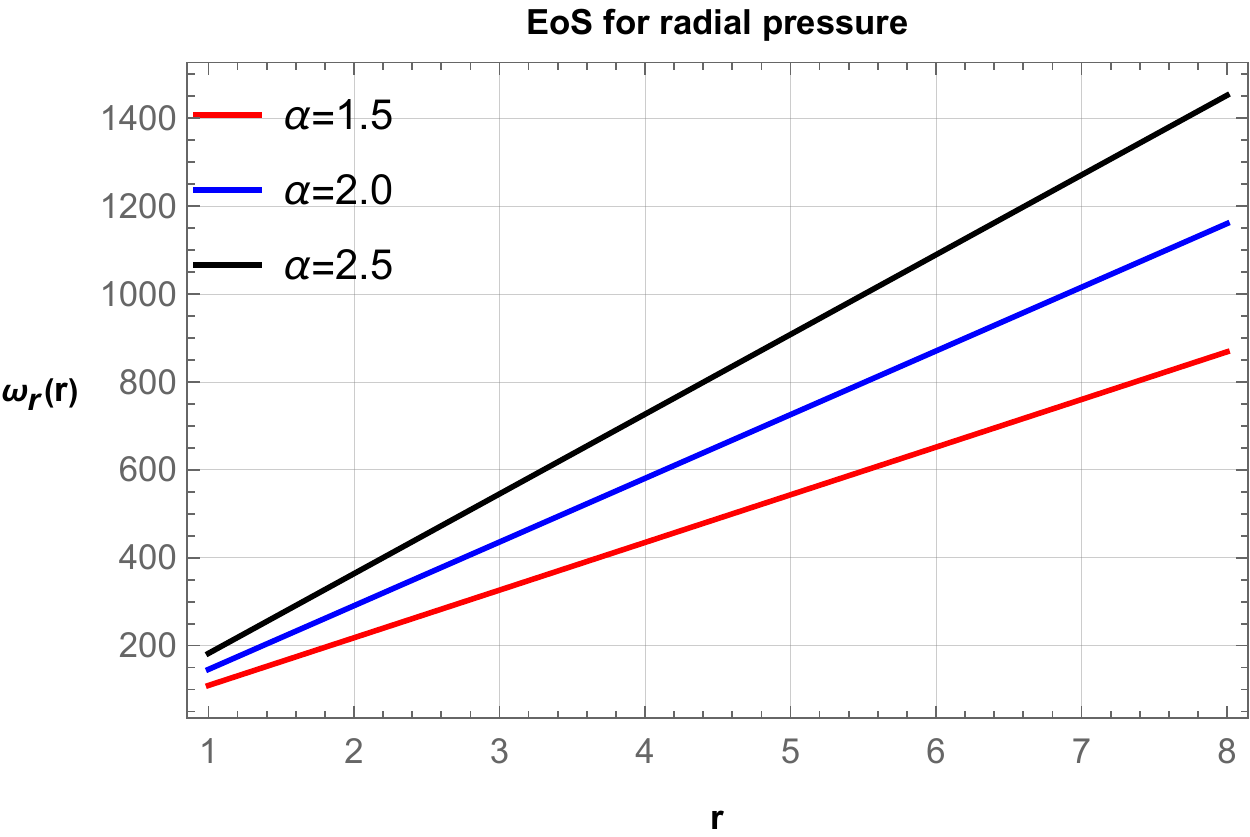}
    \includegraphics[scale=0.6]{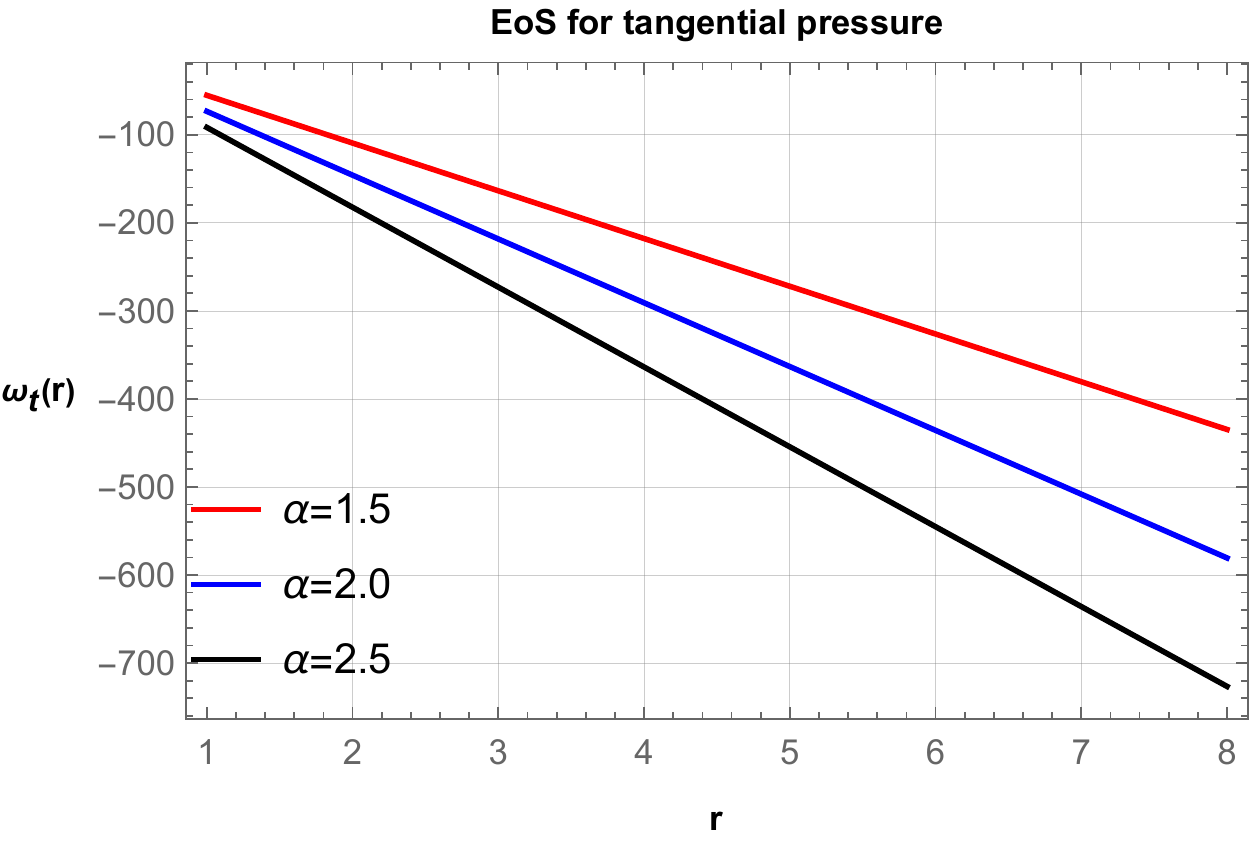}
    \caption{The EoS parameter $\omega$ for parallel plates with $r_0=1$.}
    \label{fig:6b}
\end{figure}
Fig. \ref{fig:6b} shows the behavior of EoS parameter $\omega$ against the radial distance $r$ for different values of $\alpha$. It can be observed that the radial EoS parameter increases with the increase of radial distance $r$, whereas the tangential EoS parameter decreases with the increase of radial distance.\\
Now, to study the behavior of the energy conditions, we obtained the relations among the components of anisotropic energy-momentum tensors as follows:
\begin{equation}\label{4a10}
NEC1:\quad \frac{\pi ^2 (r-2 r_0)}{720 r^4 r_0}-\frac{\alpha  r_0}{r^3},
\end{equation}
\begin{equation}\label{4a11}
NEC2:\quad -\frac{\pi ^2-720 \alpha  r_0^2}{1440 r^3 r_0},
\end{equation}
\begin{equation}\label{4a12}
DEC1:\quad \frac{\alpha  r_0}{r^3}-\frac{\pi ^2}{720 r^3 r_0},
\end{equation}
\begin{equation}\label{4a13}
DEC2:\quad \frac{\pi ^2 (r-4 r_0)-720 \alpha  r r_0^2}{1440 r^4 r_0}.
\end{equation}
At the throat of the wormhole, the expression \eqref{4a10} and \eqref{4a11} reduces to
\begin{equation}\label{4a14}
    NEC1\mid_{r=r_0}:\quad -\left[\frac{\alpha }{r_0^2}+\frac{\pi ^2}{720 r_0^4}\right],
\end{equation}
\begin{equation}\label{4a15}
 NEC2\mid_{r=r_0}:\quad \frac{\alpha }{2 r_0^2}-\frac{\pi ^2}{1440 r_0^4}.
\end{equation}
Since the right-hand side of the Eq. \eqref{4a14} has a negative sign, hence for any positive $\alpha$ and vanishing $\beta$, it is obvious that NEC will be violated for radial pressure at the wormhole throat. Also, from Eq. \eqref{4a15}, it can be concluded that the NEC will be satisfied for the tangential pressure.\\
In this case, the expression for SEC will be
\begin{equation}
    SEC: 0.
    \nonumber
\end{equation}
This result aligns with work done in \cite{Hassan/2021}. In Figs. \ref{fig:6c} and \ref{fig:6d}, we have depicted the behavior of NEC and DEC graphically. We have mentioned above that NEC is violated for radial pressure for any positive $\alpha$. DEC is also violated for tangential pressure only but satisfied for radial pressure.
\begin{figure}[h]
    \centering
    \includegraphics[scale=0.6]{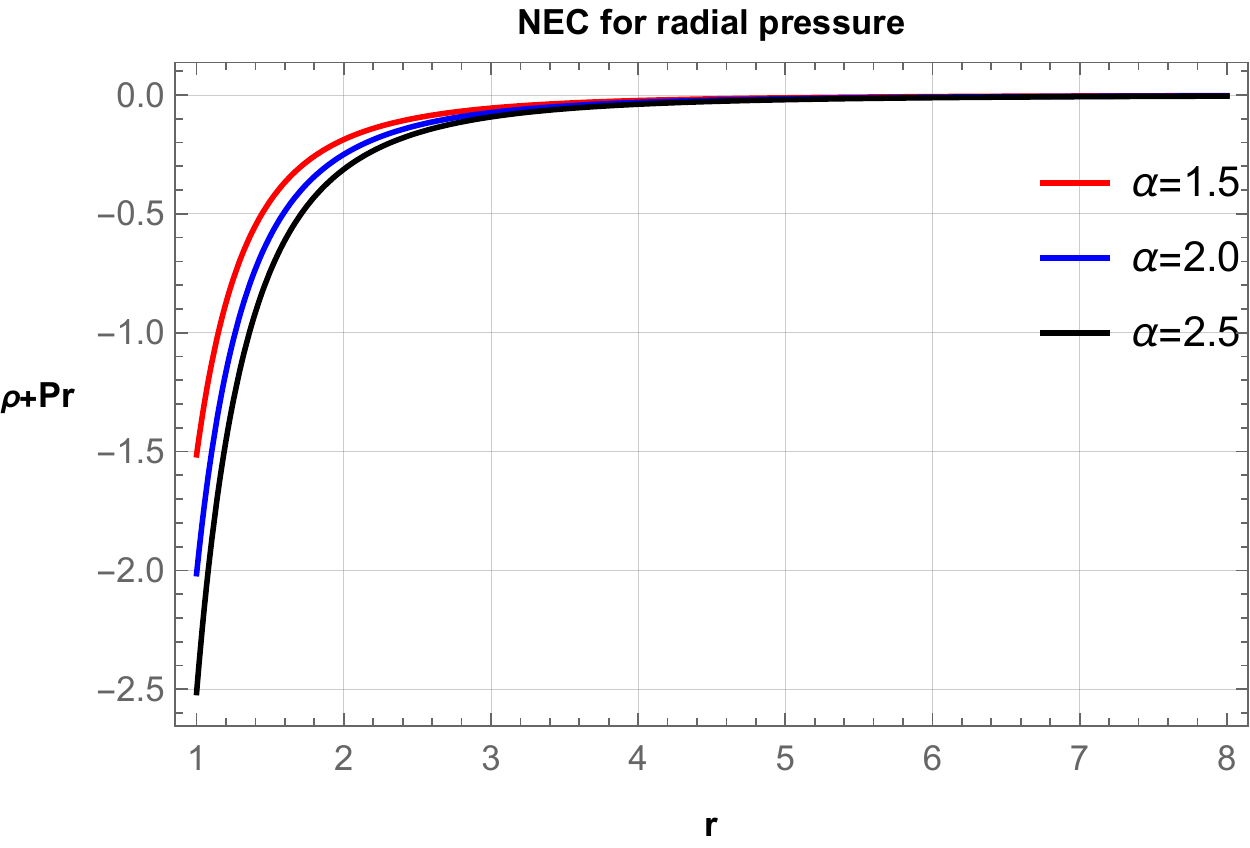}
    \includegraphics[scale=0.6]{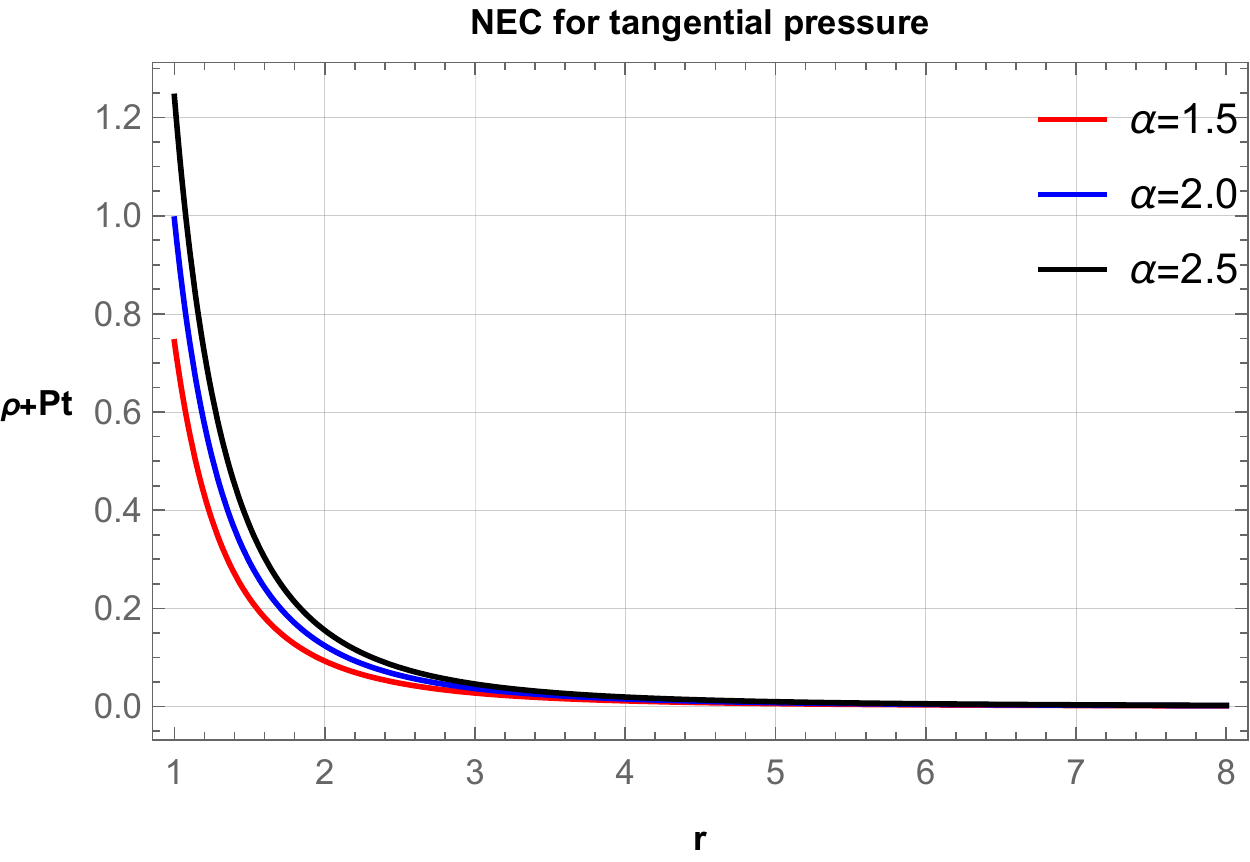}
    \caption{The variation of NEC against $r$ for parallel plates with $r_0=1$.}
    \label{fig:6c}
\end{figure}
\begin{figure}[h]
    \centering
    \includegraphics[scale=0.6]{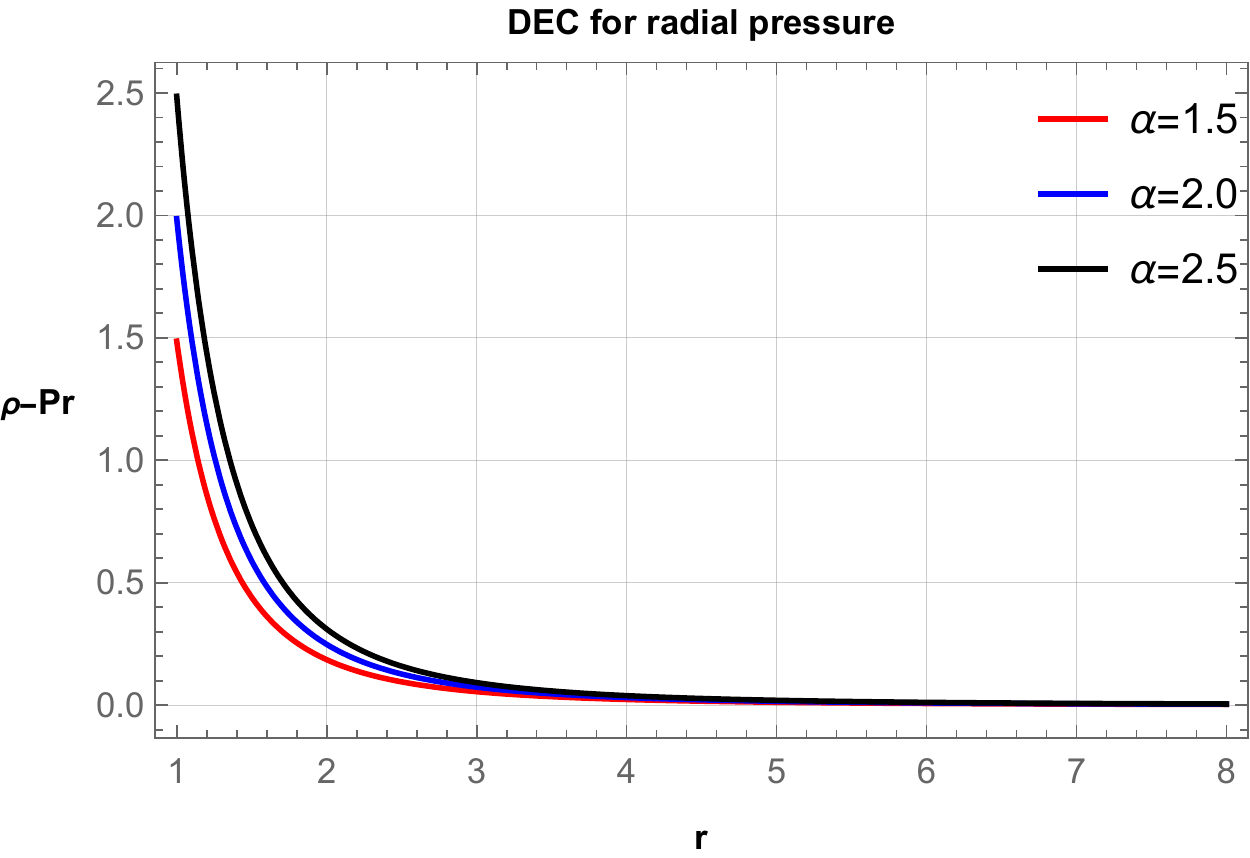}
    \includegraphics[scale=0.6]{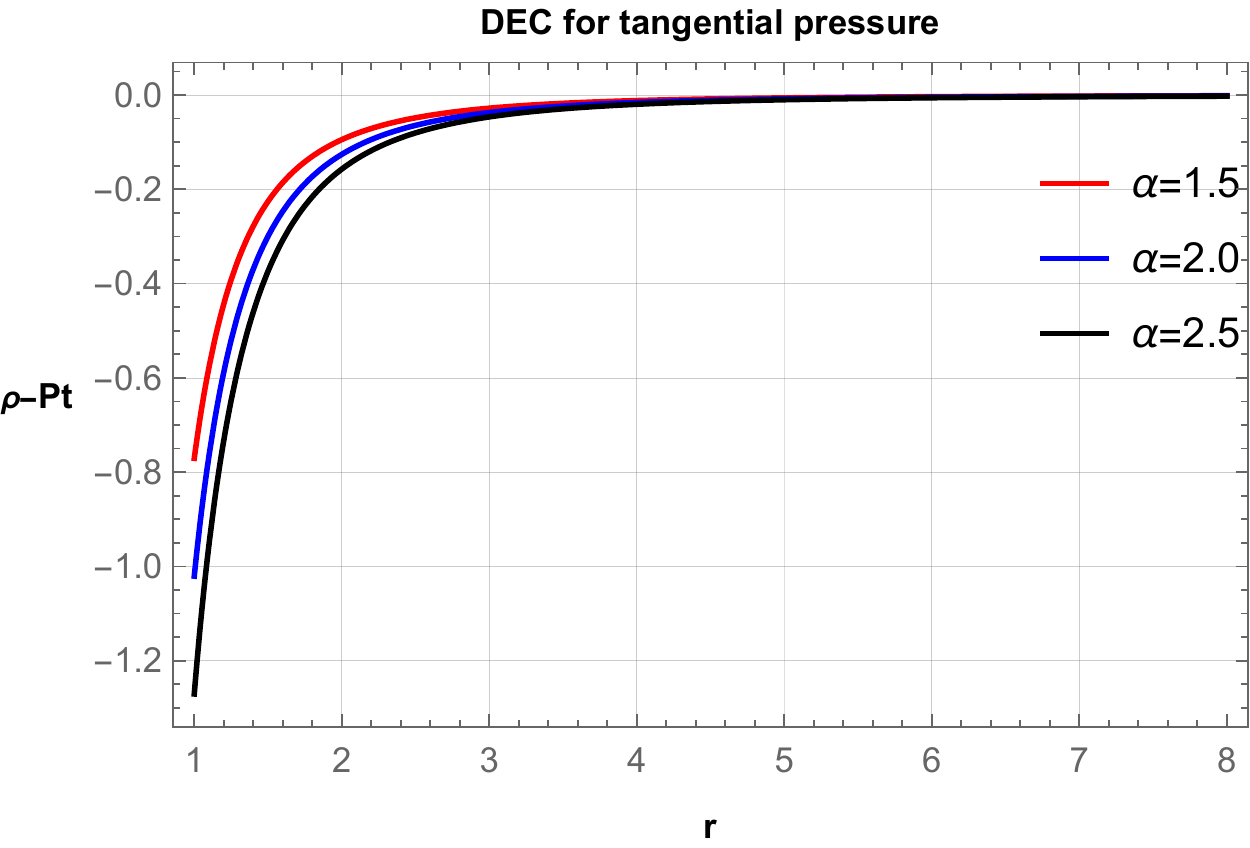}
    \caption{The variation of DEC against $r$ for parallel plates with $r_0=1$.}
    \label{fig:6d}
\end{figure}

\subsection{Case-II: Parallel Cylinder}\label{subsec2}
It is known that from \cite{cylinder,mazzitelli} (mode counting of sphere) the exact formula between Casimir interaction energy between two concentric cylinder (where length $L$ and radius $a$ and $b$, and $\gamma=\frac{b}{a}$) is given by
\begin{multline}
 E^{cc}_{12}=\frac{L}{4\pi a^2}\int_0^{\infty}d \beta \beta ln\left(\prod_{n}\left[1-\frac{I_n(\beta)K_n(\gamma \beta)}{I_n(\gamma \beta K_n{\beta})}\right]\right.\\\left.
 \times \left[1-\frac{I'_n(\beta)K'_n(\gamma \beta)}{I'_n(\gamma \beta K'_n{\beta})}\right]\right),
\end{multline}
where $I_n$ and $K_n$ are modified Bessel functions of the First kind and second kind, respectively, and $'$ denotes the derivative of the Bessel function.\\
The contribution of $I_n$ comes from the  Dirichlet boundary conditions Transverse Magnetic (TM) modes of the electromagnetic field, and the contribution of $K_n$ comes from the Neumann boundary conditions Transverse Electric (TE) modes.\\
The parallel cylinder is more feasible for practical purposes. Here, we are considering the case where two concentric cylinders of radius $a$ and $b$ (where $b>a$) and the length of both is $L$. We know that the general solutions of the energy are given by (using the PFA method)\cite{cylinder}
$$E=\frac{\pi^3L}{360a^2(\gamma-1)^3}\{1+\frac{1}{2}(\gamma-1)-(\frac{2}{\pi^2}\frac{1}{10})(\gamma-1)^2+...\}.$$
Where $\gamma$ is given by $\gamma=\frac{b}{a}$
Now we can take two limits\\
In the short distance limit ($\gamma-1<<1$) we get,
$$E=-\frac{\pi^3L}{360a^2}\frac{1}{(\gamma-1)^3}.$$
In this case we can find the $\rho$ by the fact that the volume between two concentric cylinders is $V=\pi(b^2-a^2)L$.\\
So the energy density is given by
\begin{equation}\label{4b1}
\rho=\frac{E}{\pi(b^2-a^2)L}=\frac{E}{\pi a^2(\gamma^2-1)L}.
\end{equation}
Also, the pressure is given by the Equation
$$P=-\frac{1}{S}\frac{\partial E}{\partial a}=-\frac{1}{2\pi aL}\frac{\partial E}{\partial a}=-\frac{\pi^2}{360a^4(\gamma-1)^3}$$
We have taken $b$ or the outer radius of the cylinder as our $r$ as the model. We have in our mind that the system of two parallel cylinders is inserted through the wormhole's throat. So by observing the geometry, we have taken $r$ as $b$ (outer radius).\\
Equating Eq. \eqref{14} and \eqref{4b1} and solving  the differential equation one can get the shape function as follows:
%\begin{widetext}
\begin{multline}\label{4b2}
    b(r)=-\frac{1}{17280 \alpha }\left[\frac{2 \pi ^2 \left(-2 a^2+3 a r+3 r^2\right)}{(a-r)^3}+\right.\\\left.
    \frac{3 \pi ^2 (\log (a+r)-\log (r-a))}{a}+2880 \beta  r^3\right]+c_2,
\end{multline}
%\end{widetext}
where $c_2$ is the integrating constant. Now, by imposing the throat condition $b(r_0)=r_0$ in Eq. \eqref{4b2}, we are able to find the final version of shape function $b(r)$ given by
%\begin{widetext}
\begin{multline}\label{4b3}
b(r)=r_0+\frac{1}{17280 \alpha }\left[-\frac{2 \pi ^2 \left(-2 a^2+3 a r+3 r^2\right)}{(a-r)^3}-\mathcal{F}_1\right.\\\left.
+\frac{3 \pi ^2 (\log (a+r_0)-\log (r_0-a))}{a}+2880 \beta (r_0^3-r^3)\right],
\end{multline}
where,
\begin{multline}
\mathcal{F}_1=\frac{3 \pi ^2 (\log (a+r)-\log (r-a))}{a}\\
-\frac{2 \pi ^2 \left(-2 a^2+3 a r_0+3 r_0^2\right)}{(a-r_0)^3}.
\end{multline}
%\end{widetext}

\begin{figure}[h]
\includegraphics[scale=0.6]{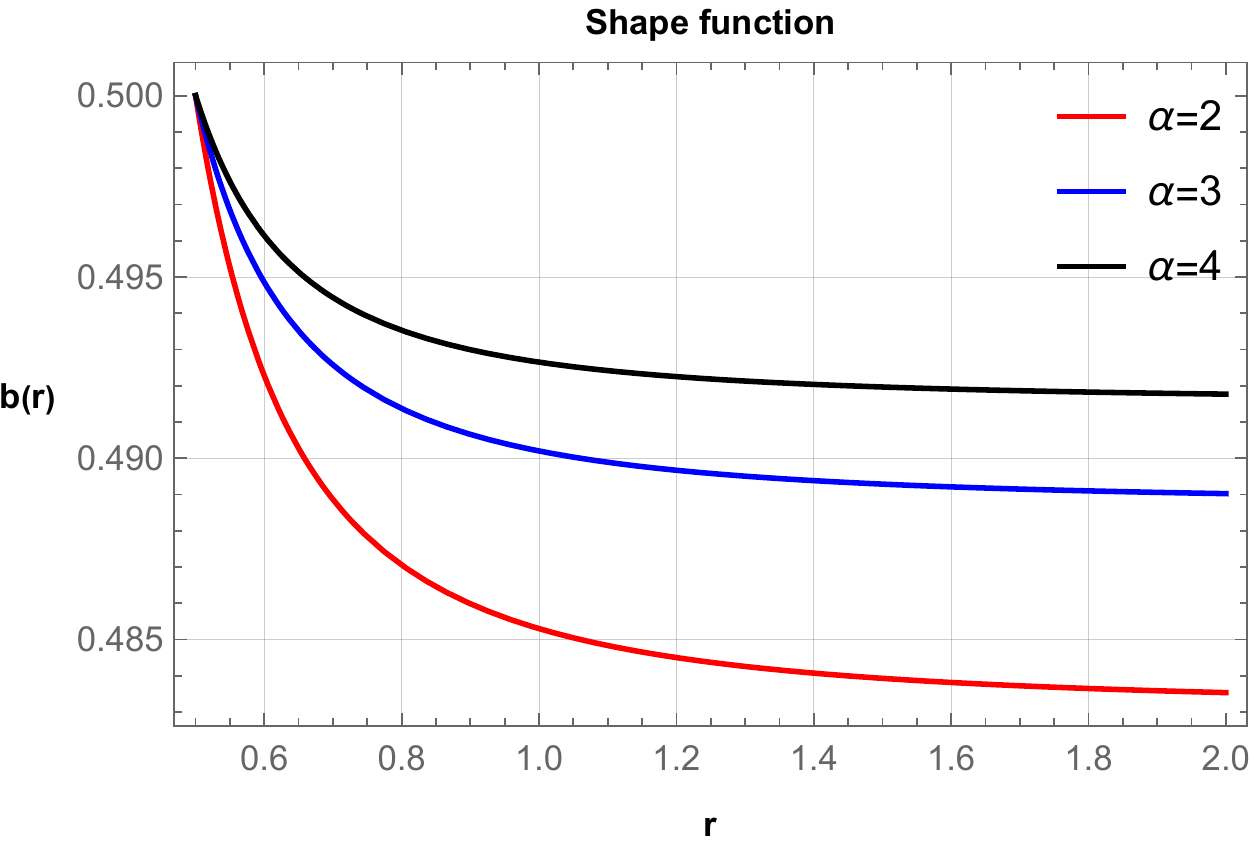}
\includegraphics[scale=0.6]{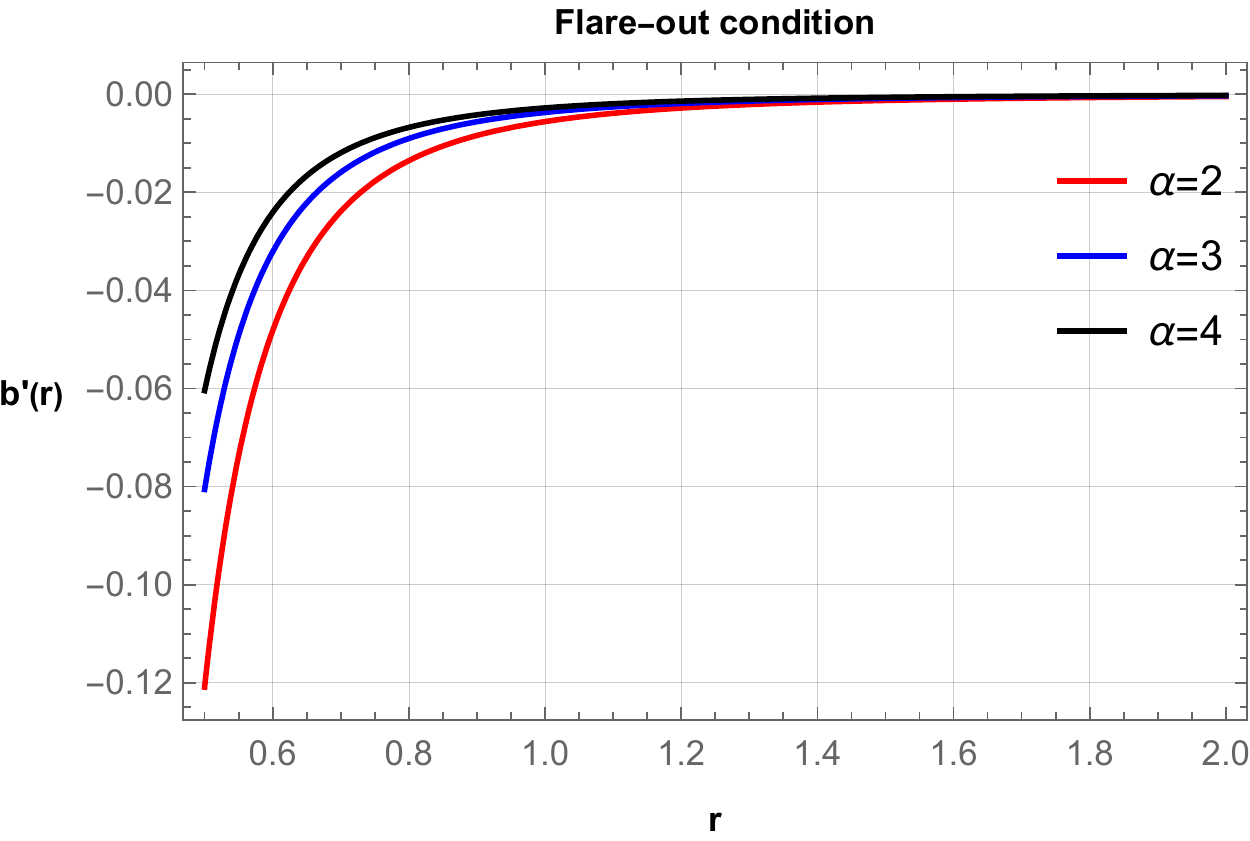}
\caption{Shape function and flare-out condition for the parallel cylinder with $r_0=0.5$.}
\label{fig6aa}
\end{figure}

\begin{figure}[h]
    \includegraphics[scale=0.6]{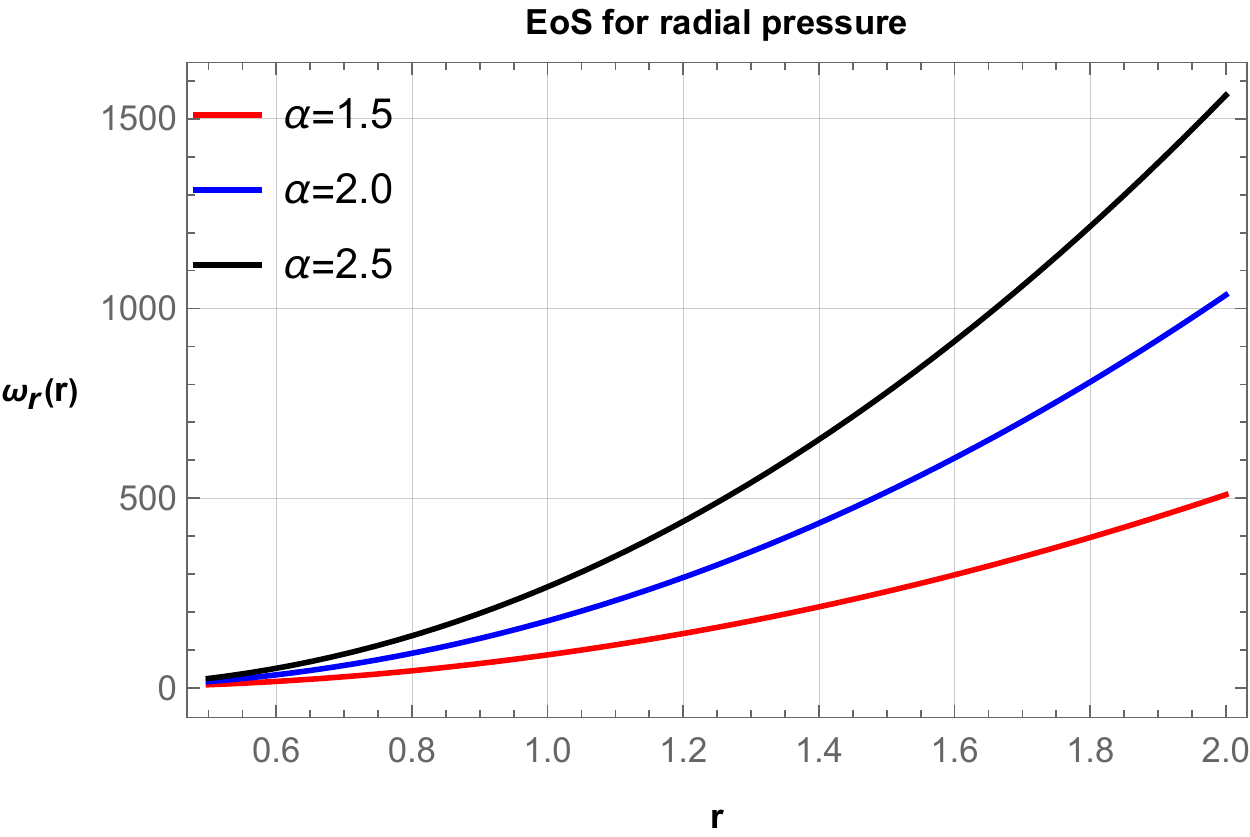}
    \includegraphics[scale=0.6]{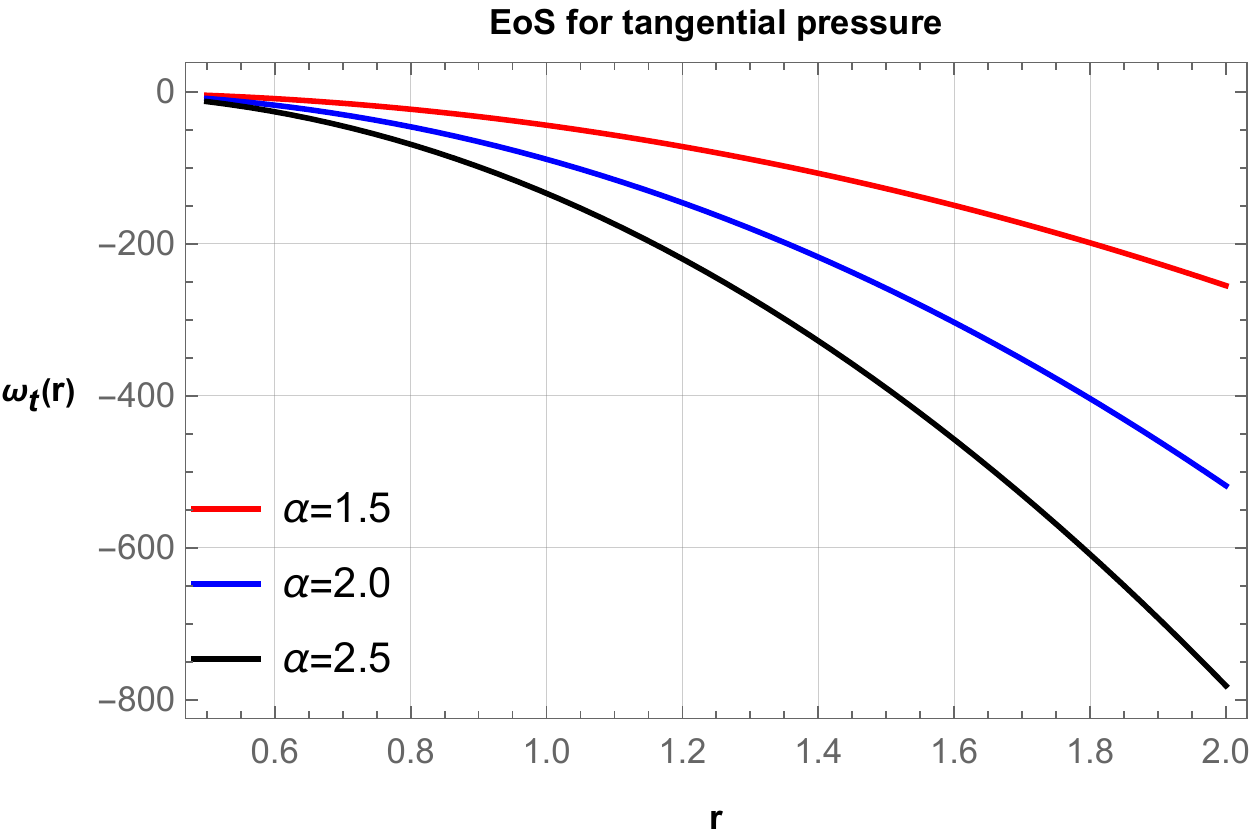}
    \caption{The EoS parameter $\omega$ for parallel cylinder with $r_0=0.5$.}
    \label{fig:41}
\end{figure}

\begin{figure}[h]
    \includegraphics[scale=0.6]{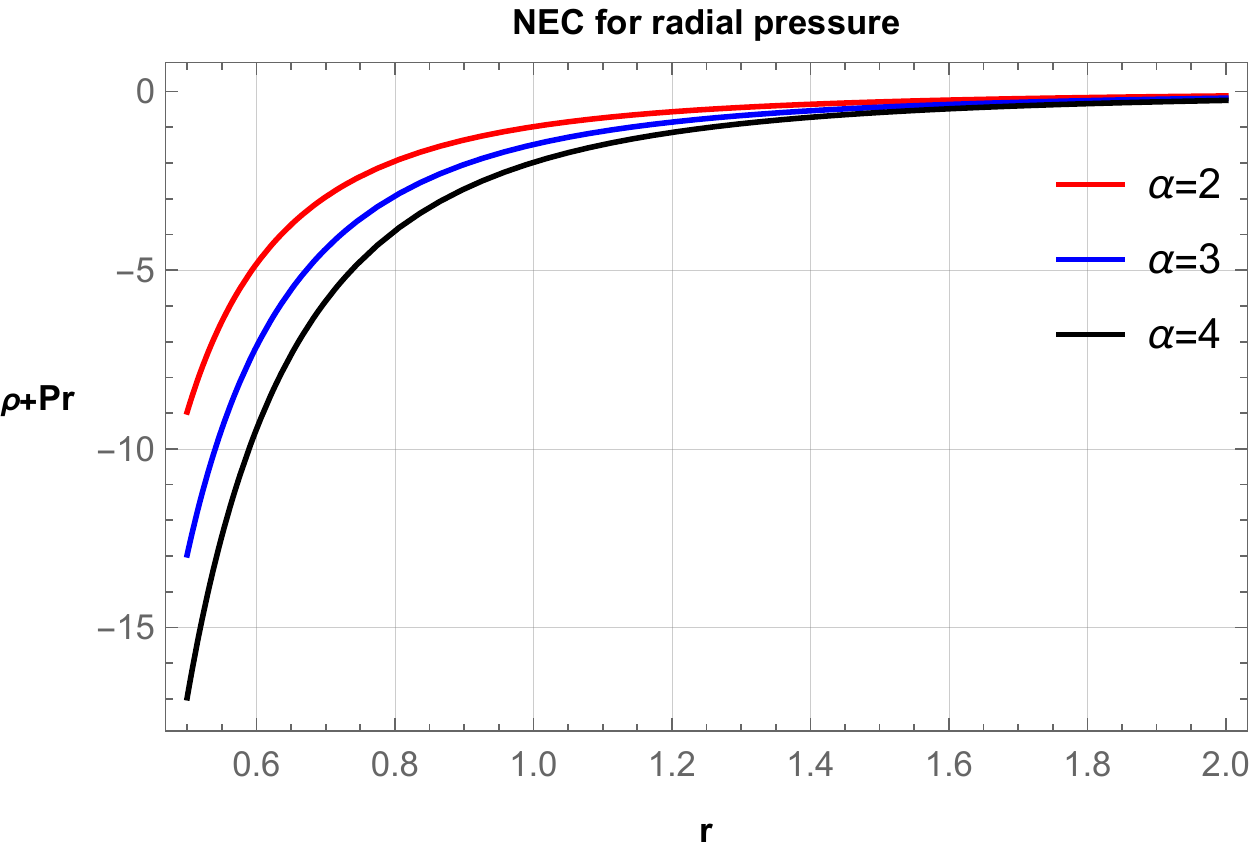}
    \includegraphics[scale=0.6]{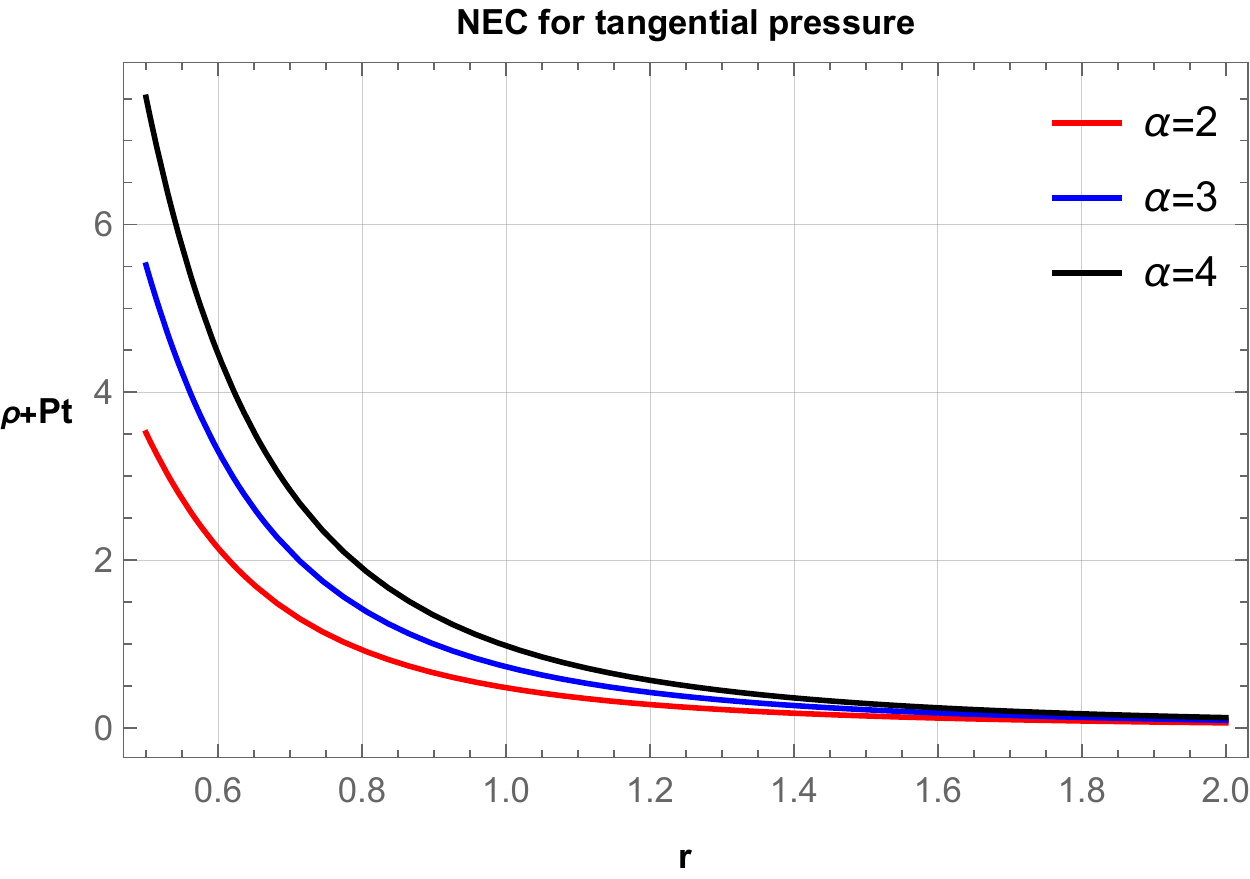}
    \caption{The variation of NEC against $r$ for parallel cylinder with $r_0=0.5$.}
    \label{fig:42}
\end{figure}

\begin{figure}[h]
    \includegraphics[scale=0.6]{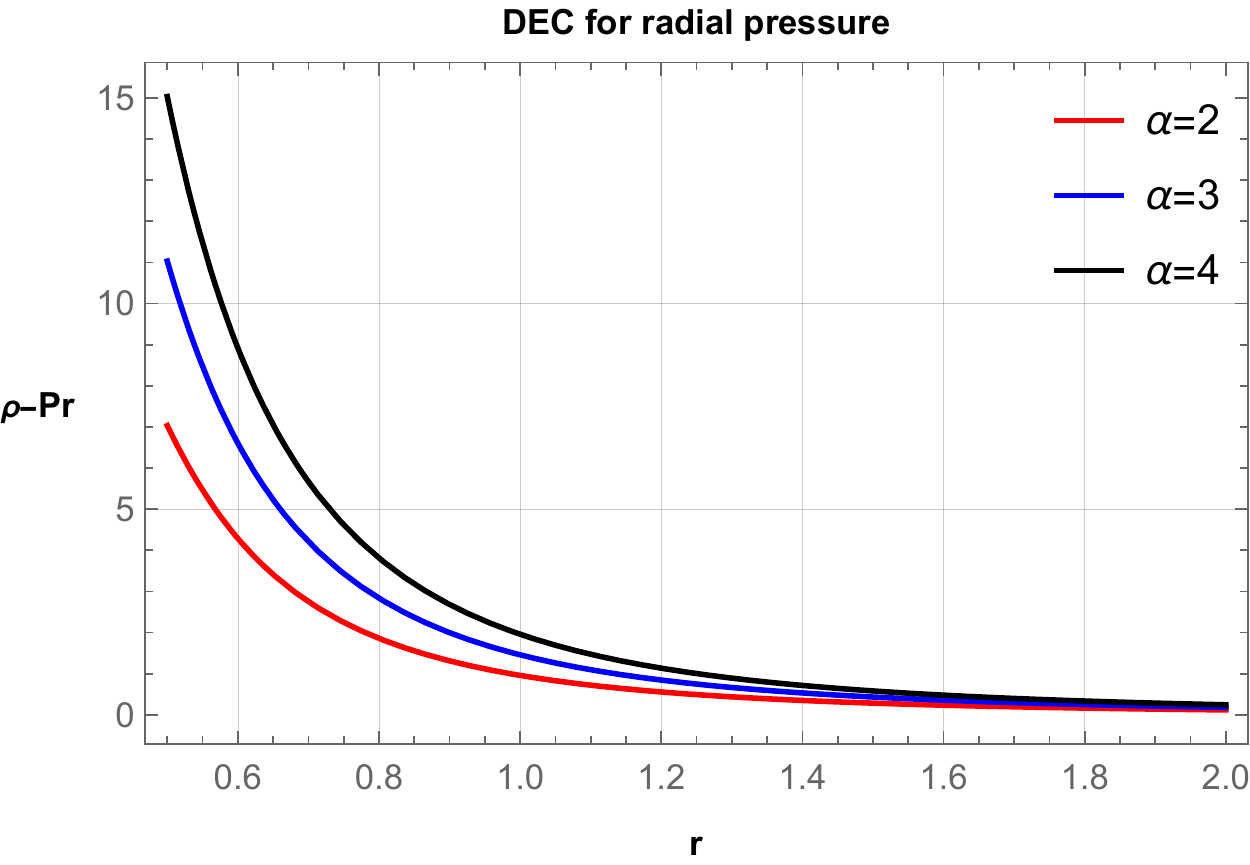}
    \includegraphics[scale=0.6]{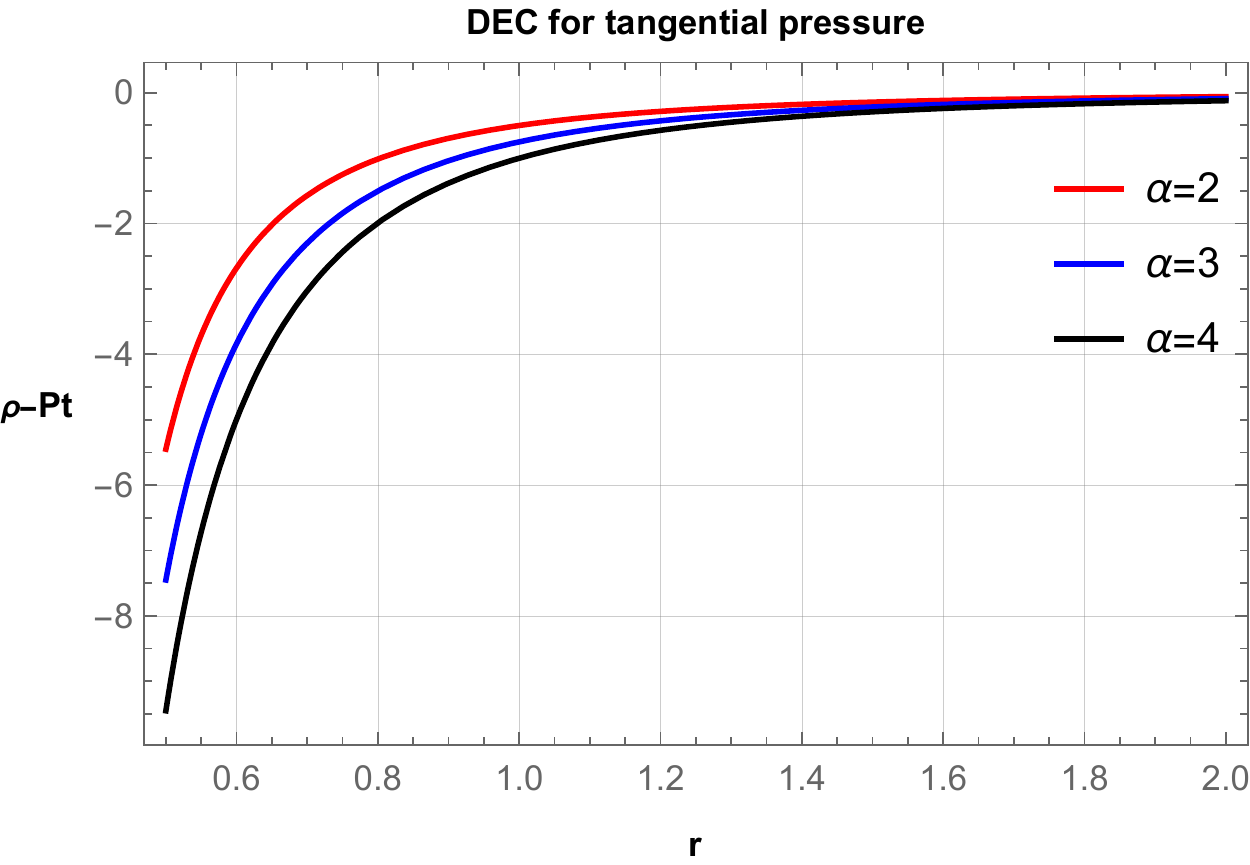}
    \caption{The variation of DEC against $r$ for parallel cylinder with $r_0=0.5$.}
    \label{fig:43}
\end{figure}

%\begin{multline}
%c_2=r_0+\frac{1}{17280 \alpha }\left[\frac{2 \pi ^2 \left(-2 a^2+3 a r_0+3 r_0^2\right)}{(a-r_0)^3}\right.\\\left.
%+\frac{3 \pi ^2 (\log (a+r_0)-\log (r_0-a))}{a}+2880 \beta  r_0^3\right]
%\end{multline}
It is noted that the obtained shape function \eqref{4b3} does not follow the asymptotically flatness condition, which is a necessary condition of the shape function. We noticed that for vanishing $\beta$, this condition would satisfy. Therefore, for $\beta\rightarrow 0$, the Eq. \eqref{4b3} reduces to 
\begin{multline}\label{4b33}
b(r)=r_0+\frac{1}{17280 \alpha }\left[-\frac{2 \pi ^2 \left(-2 a^2+3 a r+3 r^2\right)}{(a-r)^3}-\mathcal{F}_1\right.\\\left.
+\frac{3 \pi ^2 (\log (a+r_0)-\log (r_0-a))}{a}\right],
\end{multline}
 The graphical behavior of the shape functions has been shown in Fig. \ref{fig6aa}. One can observe that $b(r)$ shows positively decreasing behavior for any positive values of $\alpha$. Also, the flaring-out condition is satisfied in the vicinity of the throat under asymptotic background. In this case, we consider the throat radius $r_0=0.5$.
 %We also observed that for any non-zero $\beta$, the asymptotically flatness condition would not be satisfied.
For the parallel cylinder case, the components of energy-momentum tensors define as
\begin{equation}\label{4b4}
\rho=-\frac{\pi ^2 a}{360 (a-r)^4 (a+r)},
\end{equation}
\begin{multline}\label{4b5}
P_r=\frac{1}{17280 r^3}\left[\frac{2 \pi ^2 \left(-2 a^2+3 a r+3 r^2\right)}{(a-r)^3}+\mathcal{F}_1\right.\\\left.
-\frac{3 \pi ^2 (\log (a+r_0)-\log (r_0-a))}{a}\right]-\frac{\alpha  r_0}{r^3},
\end{multline}
\begin{multline}\label{4b6}
P_t=\frac{1}{34560}\left[\frac{1}{r^3}\left(-\frac{2 \pi ^2 \left(-2 a^2+3 a r+3 r^2\right)}{(a-r)^3}\right.\right.\\\left.\left.
-\mathcal{F}_1+\frac{3\pi^2}{a}\left(\log (a+r_0)
-\log (r_0-a)\right)\right)\right.\\\left.
+\frac{48 \pi ^2 a}{(a-r)^4 (a+r)}
+\frac{17280 \alpha  r_0}{r^3}\right].
\end{multline}
Taking the Eqs. (\ref{4b4}-\ref{4b6}) into account, we have plotted the graphs for EoS parameter $\omega(r)$ in Fig. \ref{fig:41}. As usual, we could observe that the radial EoS parameter increases with the radial distance, whereas tangential EoS parameter shows completely opposite behavior, i.e., it decreases with the increasing of radial distance $r$.\\
Let us now discuss the energy conditions for the parallel cylinder case. Considering the Eqs. (\ref{4b4}-\ref{4b6}), we can write the expressions for NEC as follows:
\begin{multline}\label{4b7}
NEC1:\quad \frac{1}{17280}\left[\frac{1}{r^3}\left(\frac{2 \pi ^2 \left(-2 a^2+3 a r+3 r^2\right)}{(a-r)^3}\right.\right.\\\left.\left.
+\mathcal{F}_1-\frac{1}{a}\left(3 \pi ^2 (\log (a+r_0)-\log (r_0-a))\right)\right)\right.\\\left.
-\frac{48 \pi ^2 a}{(a-r)^4 (a+r)}-\frac{17280 \alpha  r_0}{r^3}\right],
\end{multline}
\begin{multline}\label{4b8}
NEC2:\quad \frac{1}{34560 r^3 (a-r)^4}\left[\frac{1}{a (a-r_0)^3}\left((a-r)\right.\right.\\\left.\left.
\times \left((a-r_0)^3 \left(2 \pi ^2 a \left(2 a^2-3 a r-3 r^2\right)\right.\right.\right.\right.\\\left.\left.\left.\left.
+3 \pi ^2 (a-r)^3 (\log (r-a)-\log (a+r))\right)\right.\right.\right.\\\left.\left.\left.
-(a-r)^3 \left(2 \pi ^2 a \left(2 a^2-3 a r_0-3 r_0^2\right)\right.\right.\right.\right.\\\left.\left.\left.\left.
+3 \pi ^2 (a-r_0)^3 (\log (r_0-a)-\log (a+r_0))\right)\right.\right.\right.\\\left.\left.\left.
+17280 a \alpha  r_0 (a-r)^3 (a-r_0)^3\right)\right)-\frac{48 r^3\pi ^2 a}{a+r}\right].
\end{multline}
At the throat of the wormhole, i.e., at $r=r_0$, the radial NEC reduces to
\begin{equation}\label{4b9}
NEC1\mid_{r=r_0}=-\left(\frac{\pi ^2 a}{360 (a-r_0)^4 (a+r_0)}+\frac{\alpha }{r_0^2}\right).
\end{equation}
It can be observed from the above expression that NEC is violated for radial pressure at the wormhole throat as the RHS of Eq. \eqref{4b9} is a negative quantity. One can check the graphical behavior of NEC for both pressures, which is depicted in Fig. \ref{fig:42}. Moreover, we have studied the DEC, which has shown in Fig. \ref{fig:43}. We have observed that for any positive $\alpha$, radial DEC is satisfied, whereas tangential DEC is violated in the entire spacetime. Further, we checked SEC, which vanishes in the entire spacetime. This similar situation exactly matches the previous case.

\subsection{Case-III: Two Spheres}\label{subsec3}
We can also calculate the energy between two spheres, both having radius $A$, and $ D$ gives the distance between them. We also assume that $D>>A$ using the multiple scattering approach given in \cite{Sphere} also checking with the consistency \cite{kardar} and \cite{klich}.\\
We also note that for engineering purposes, one can estimate radius ($A$) and distance ($D$) from the embedding picture of the wormhole, which can be seen in \cite{Channuie}.\\
We can use an equatorial slice at $\theta=\frac{\pi}{2}$ and also at a fixed moment in time to find the metric as
\begin{equation}\label{12345}
    ds^2=\frac{dr^2}{1-\frac{b(r)}{r}}+r^2d\phi^2.
\end{equation}
Now, if we want to embed the above metric in a cylindrical coordinate, we can get
\begin{equation}\label{123456}
    ds^2=dz^2+dr^2+r^2d\phi^2.
\end{equation}
comparing the above expressions, Eqs. \eqref{12345} and \eqref{123456}, we can find
\begin{equation}\label{1234567}
    \frac{dz}{dr}=\pm\sqrt{\frac{r}{r-b(r)}-1}.
\end{equation}
As we can see from the above expression, we have found the metric of wormhole in Euclidian plane, so given shape function $b(r)$ one can estimate distance ($D$), noting that $D\approx 2z$ ($z$ can be found from Eq. \eqref{1234567}) and radius ($A$) would be $A\approx r$.\\
We use the formula given in \cite{Sphere} via multiple scattering approach(which is only valid as long as $D>>A$).
\begin{equation}
E=-\frac{1}{8\pi D}\left[ln(\frac{D^2}{D^2-4A^2})+\frac{40A^4-6A^2D^2}{(D^2-4A^2)^2}+\frac{2A^2}{D^2}\right].
\nonumber
\end{equation}
In the case of two spheres, we consider the radius of the sphere $A$ as the radius $r$ of the wormhole.\\
In this case, we can find the $\rho$ by the fact that the volume between two spheres is $V=\pi A^2D$ (assuming $D>>A$). So the energy density is given by
\begin{equation}\label{4c1}
 \rho=\frac{E}{\pi A^2D}   .
\end{equation}
Also, the pressure define by
\begin{equation}\label{4c2}
    P=-\frac{1}{S}\frac{\partial E}{\partial D}.
\end{equation}
On solving the above equation, we get
\begin{multline}
P=-\frac{1}{2\pi A^2}\left[\frac{1}{8\pi D^2}\left(ln(\frac{D^2}{D^2-4A^2})+\frac{40A^4-6A^2D^2}{(D^2-4A^2)^2}\right.\right.\\\left.\left.
+\frac{2A^2}{D^2}\right)-\frac{1}{8\pi D}\left(\frac{2}{D}-\frac{2D}{D^2-4A^2}-\frac{12A^2D}{(D^2-4A^2)^2}\right.\right.\\\left.\left.
-\frac{160A^4D-24A^2D^3}{(D^2-4A^2)^3}-\frac{4A^2}{D^3}\right)\right].
\end{multline}

One can notice that for this two-sphere case, we have two dependent variables, $A$ and $D$. Hence, in the following consecutive subsections, we shall investigate the effect of this Casimir source in wormhole geometry by putting some restrictions on $A$ and $D$.

\subsubsection{$D$ variable and $A$  constant}
\begin{figure}[h]
\includegraphics[scale=0.6]{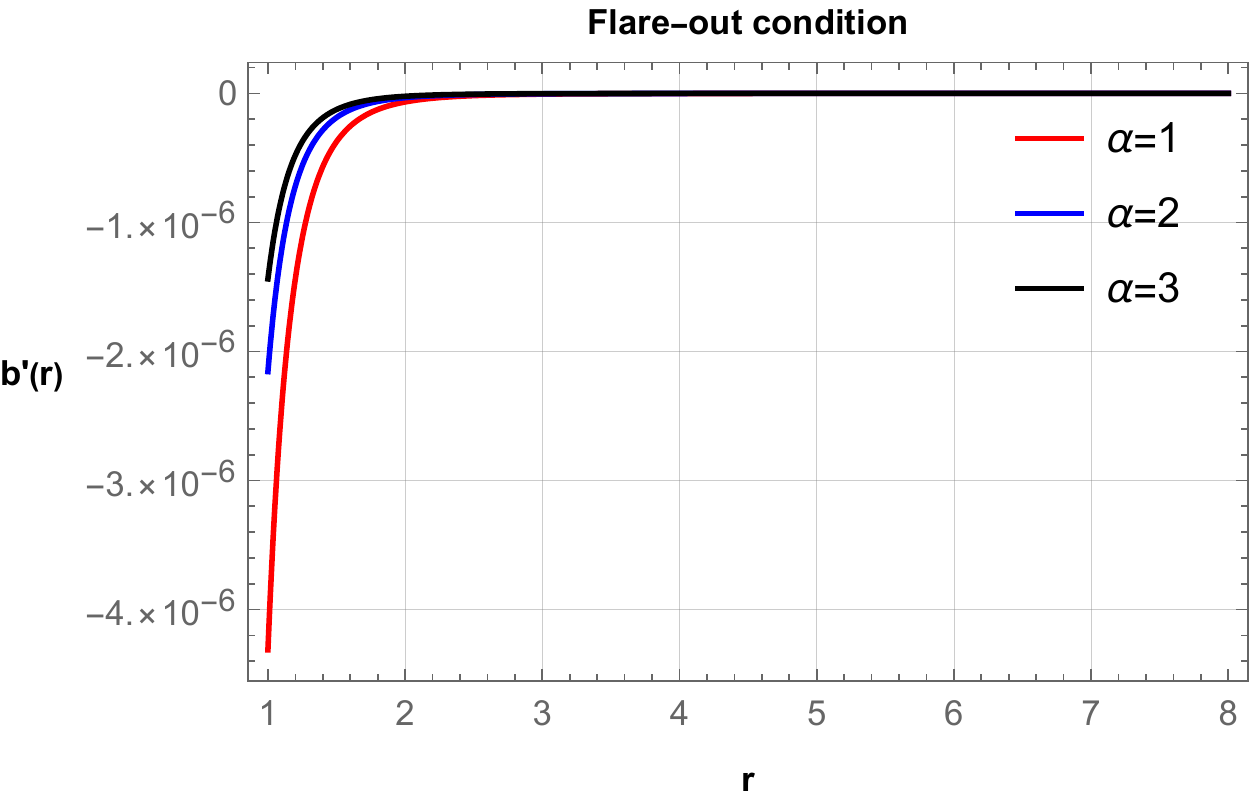}
\caption{Flare-out condition for two-sphere (case-1) with with $r_0=1$ and $A=0.05$.}
\label{fig6aaaa}
\end{figure}

\begin{figure}[h]
    \includegraphics[scale=0.6]{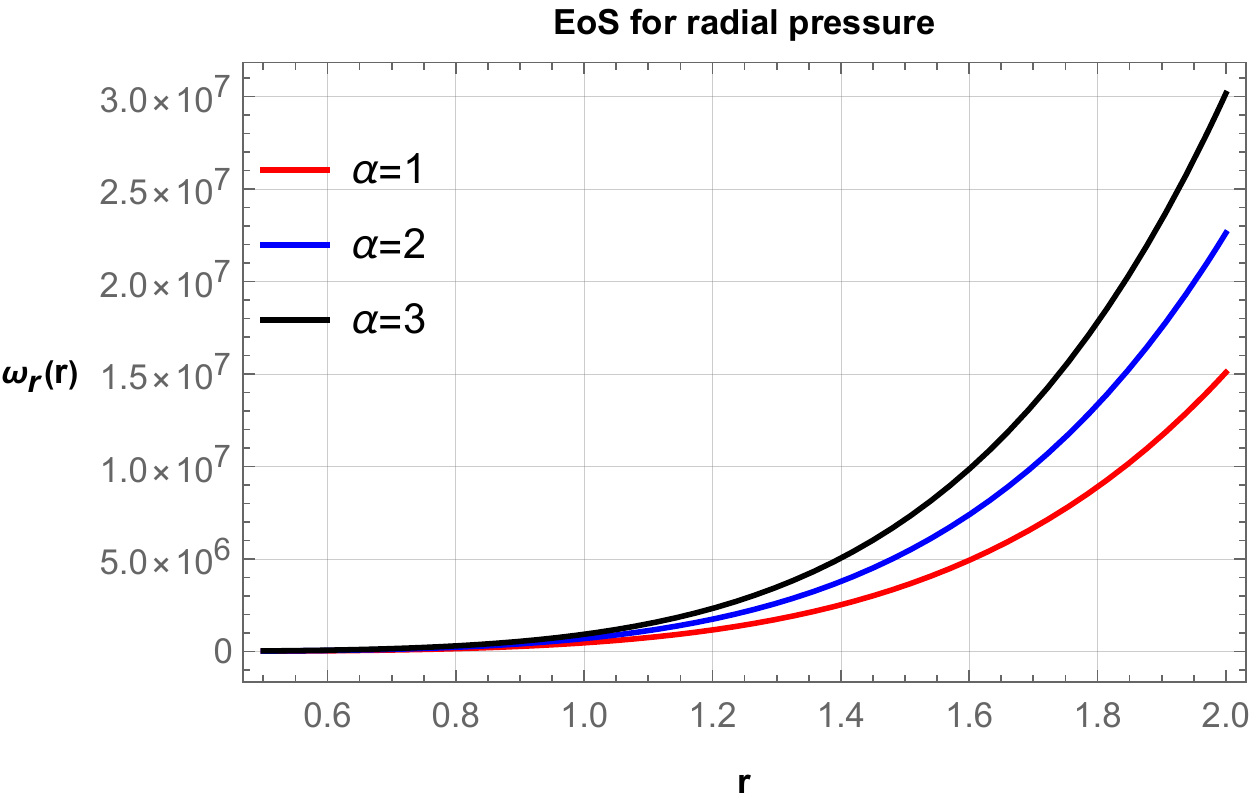}
    \includegraphics[scale=0.6]{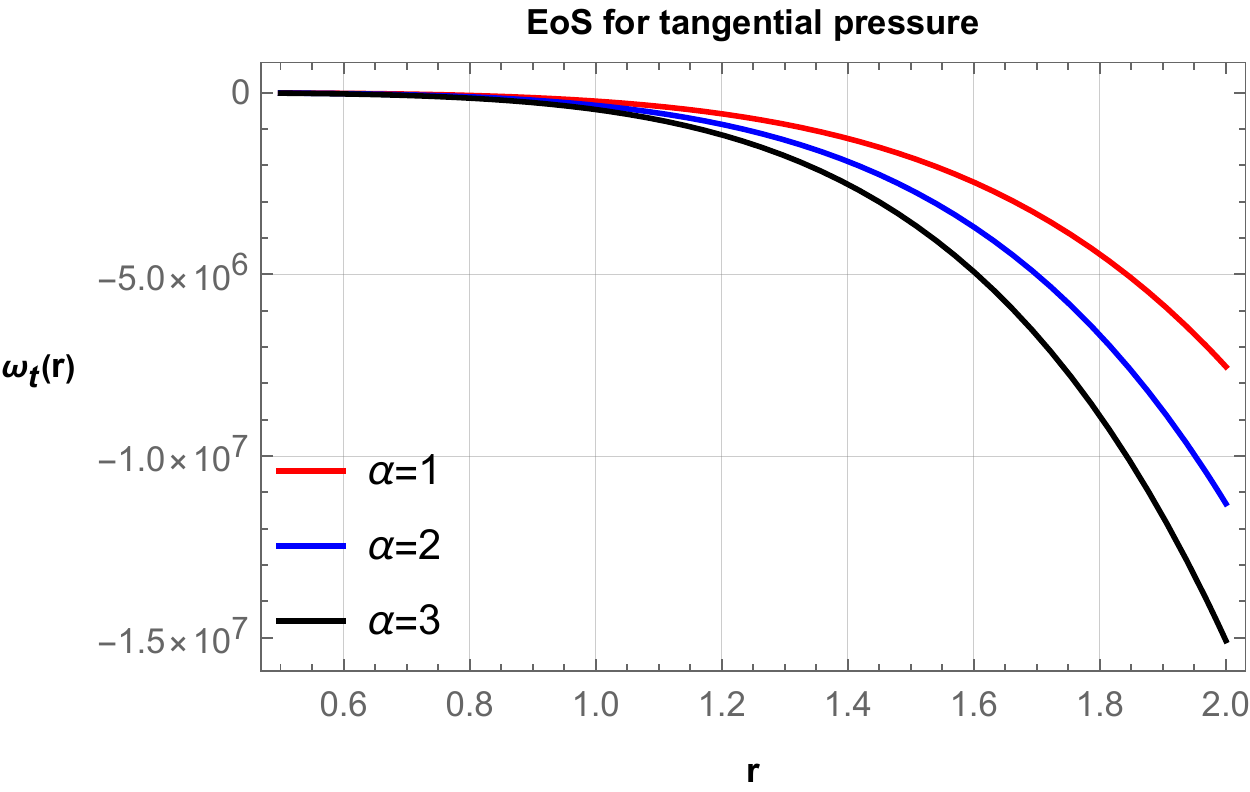}
    \caption{ The EoS parameter $\omega$ for two-sphere (case-1) with $r_0=1$ and $A=0.05$.}
    \label{fig:61}
\end{figure}

\begin{figure}[h]
    \includegraphics[scale=0.6]{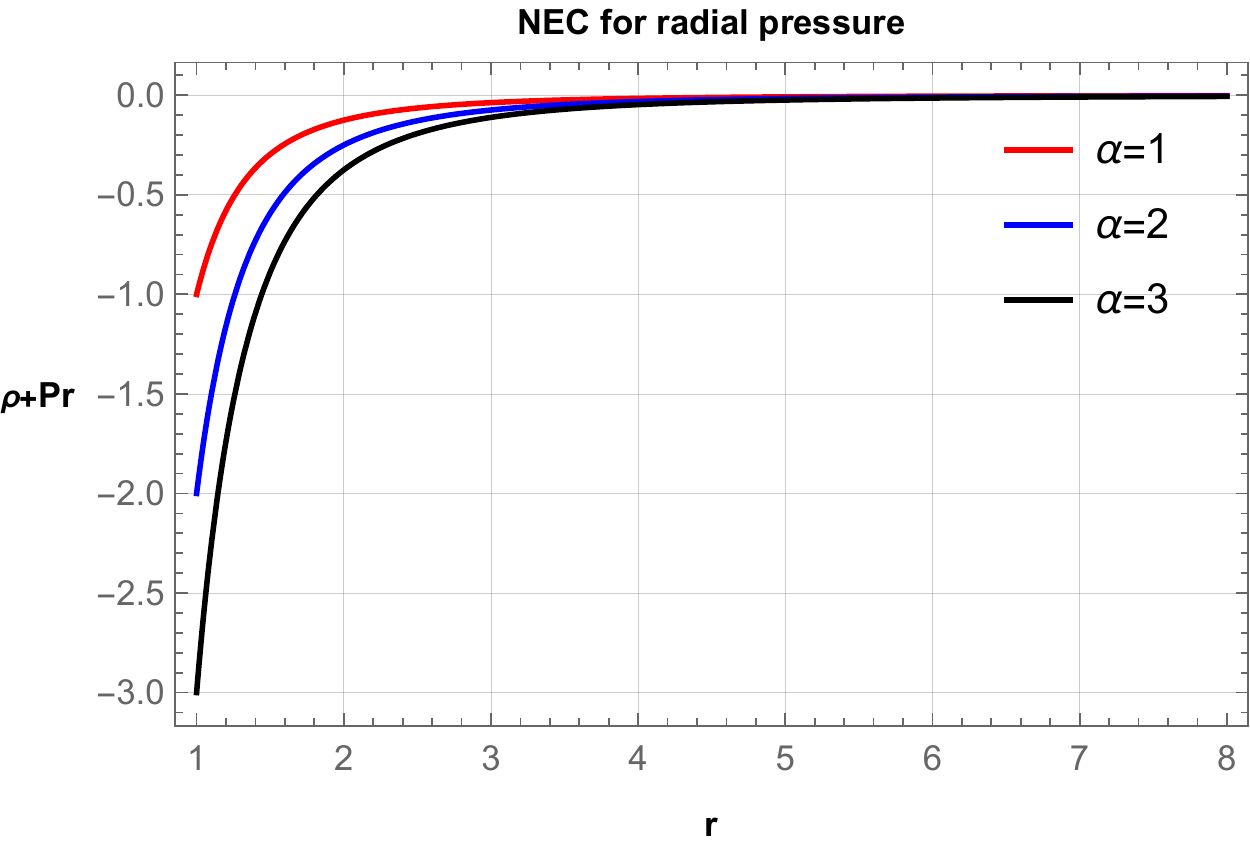}
    \includegraphics[scale=0.6]{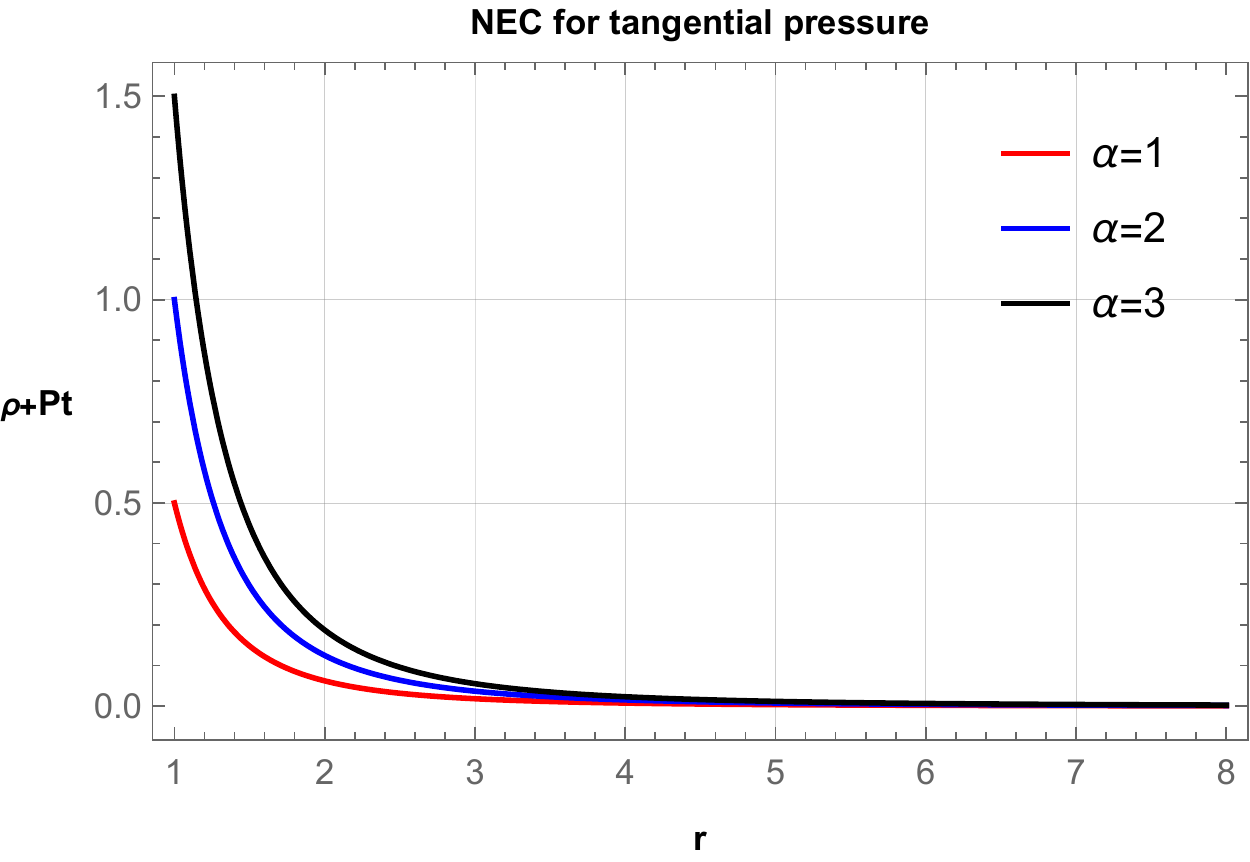}
    \caption{The variation of NEC against $r$ for two sphere (case-1) with $r_0=1$ and $A=0.05$.}
    \label{fig:5=62}
\end{figure}

\begin{figure}[h]
    \includegraphics[scale=0.6]{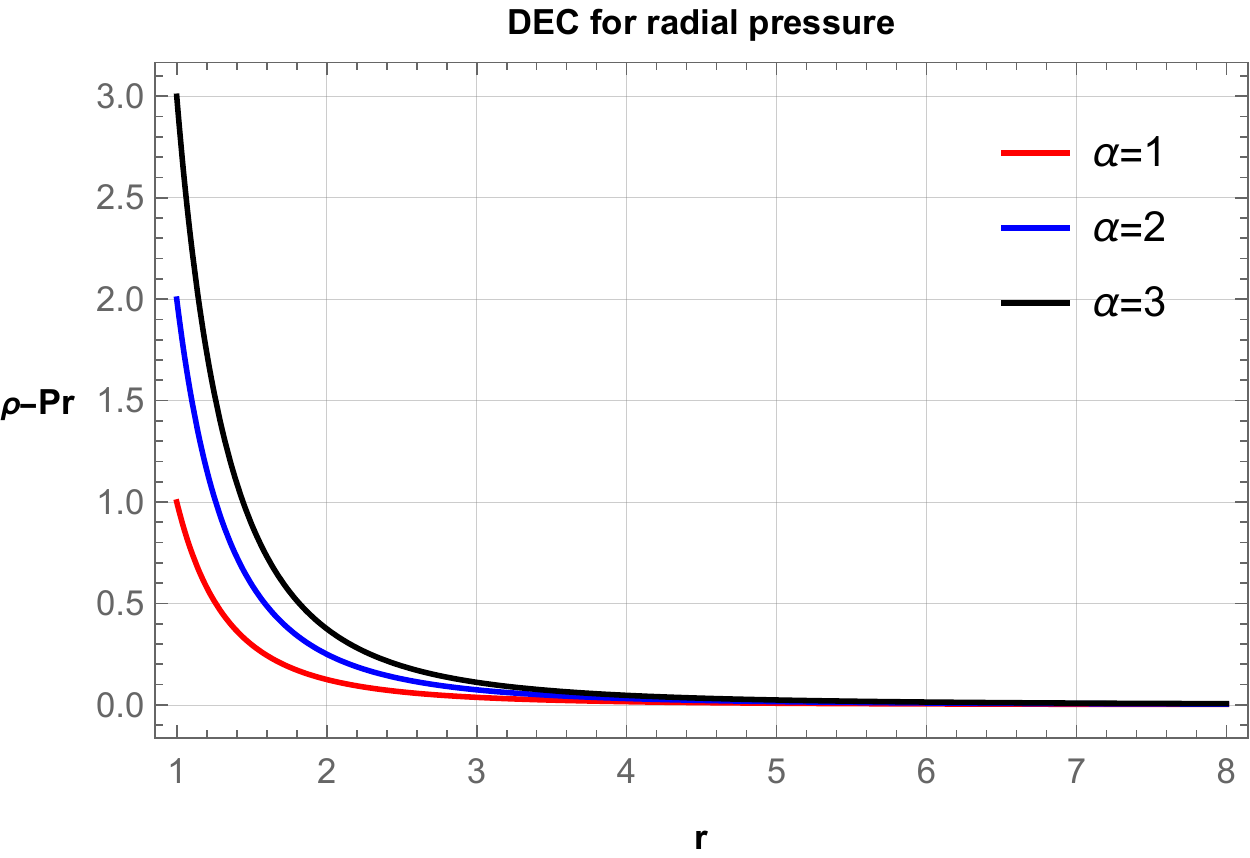}
    \includegraphics[scale=0.6]{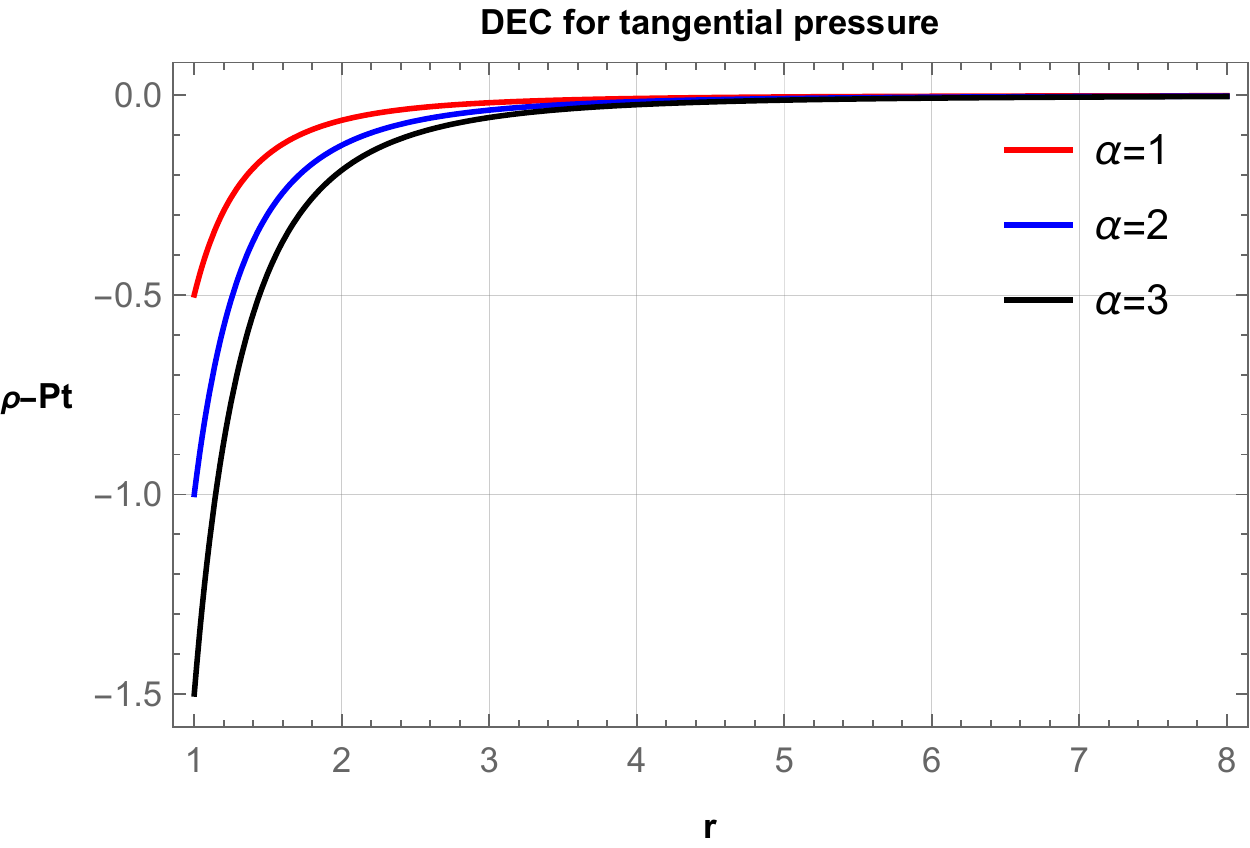}
    \caption{The variation of DEC against $r$ for two sphere (case-1) with with $r_0=1$ and $A=0.05$.}
    \label{fig:63}
\end{figure}

For this case, we replace $D$ by wormhole radius $r$ and considered the sphere radius $A$ as constant. The shape function can be obtained by comparing the Eqs. \eqref{14} and \eqref{4c1}
\begin{multline}
b(r)=-\frac{1}{8 \pi ^2 \alpha  A^2}\left[\frac{4}{3} \pi ^2 A^2 \beta  r^3-\frac{2 A^2 r}{r^2-4 A^2}\right.\\\left.
+r \log \left(\frac{r^2}{r^2-4 A^2}\right)-\frac{2 A^2}{r}\right]+c_3,
\end{multline}
where $c_3$ is the integrating constant. By imposing the throat condition on the above expression, we could able to find the shape function $b(r)$ as follows
\begin{multline}\label{4cc1}
b(r)=-\frac{1}{8 \pi ^2 \alpha  A^2}\left[\frac{2 A^2 \left(-\frac{3 r^2}{r^2-4 A^2}+2 \pi ^2 \beta  r^4-3\right)}{3 r}\right.\\\left.
+r \log \left(\frac{r^2}{r^2-4 A^2}\right)\right]+\frac{1}{8 \pi ^2 \alpha  A^2}\left[r_0 \log \left(\frac{r_0^2}{r_0^2-4 A^2}\right)\right.\\\left.
+\frac{2 A^2 \left(-\frac{3 r_0^2}{r_0^2-4 A^2}+2 \pi ^2 \beta  r_0^4-3\right)}{3 r_0}\right]+r_0.
\end{multline}
The above expression \eqref{4cc1} is not asymptotically flat. However, we notice that for any $\alpha$ and vanishing $\beta$, the asymptotically flatness condition will satisfy. Thus, the final shape function $b(r)$ for this case can be read as
\begin{multline}\label{4ccc1}
b(r)=r_0-\frac{1}{8 \pi ^2 \alpha  A^2}\left[\frac{2 A^2 \left(-\frac{3 r^2}{r^2-4 A^2}-3\right)}{3 r}\right.\\\left.
+r \log \left(\frac{r^2}{r^2-4 A^2}\right)\right]+\frac{1}{8 \pi ^2 \alpha  A^2}\left[r_0 \log \left(\frac{r_0^2}{r_0^2-4 A^2}\right)\right.\\\left.
+\frac{2 A^2 \left(-\frac{3 r_0^2}{r_0^2-4 A^2}-3\right)}{3 r_0}\right].
\end{multline}
Also, we observed that for any positive $\alpha$, the flaring out condition, $b^{'}(r_0)<1$, would satisfy at the throat of the wormhole. In Fig. \ref{fig6aaaa}, we have depicted the behavior of the flare-out condition.
As already defined in the previous sections, the EoS for both pressures can be read as
$$\omega_r(r)=\frac{P_r}{\rho}, \quad \omega_t(r)=\frac{P_t}{\rho}.$$
Inserting Eqs. (\ref{14}-\ref{16}) and \eqref{4ccc1} in the above expressions, we obtained the following EoS parameters for both pressures
\begin{multline}
\omega_r(r)=-\frac{1}{\mathcal{H}_1}\left[\left(4 A^2-r^2\right) \left(4 A^2 \left(-6 A^2 \left(4 \pi ^2 \alpha  r r_0^2 \right.\right.\right.\right.\\\left.\left.\left.\left.
\times \left(r^2+r_0^2\right)+2 r^2 r_0-r^3-2 r r_0^2+r_0^3\right)+24 A^4 \right.\right.\right.\\\left.\left.\left.
\times \left(r \left(4 \pi ^2 \alpha  r_0^2-1\right)+r_0\right)+3 r^2 r_0^2 \left(r \left(2 \pi ^2 \alpha  r_0^2-1\right)\right.\right.\right.\right.\\\left.\left.\left.\left.
+r_0\right)\right)+3 r r_0 \left(r^2-4 A^2\right)
\left(4 A^2-r_0^2\right) \mathcal{H}_2\right)\right],
\end{multline}
\begin{multline}
\omega_t(r)=\frac{1}{2 \mathcal{H}_1}\left[4 A^2 \left(\mathcal{H}_3-24 A^2 r^3 r_0^2+48 A^4 r^3\right.\right.\\\left.\left.
-6 A^2 r^5+384 \pi ^2 \alpha  A^6 r r_0^2-96 \pi ^2 \alpha  A^4 r r_0^4+48 A^4 r r_0^2 \right.\right.\\\left.\left.
-96 A^6 r-48 A^4 r_0^3+192 A^6 r_0-6 \pi ^2 \alpha  r^5 r_0^4 \right.\right.\\\left.\left.
+3 r^5 r_0^2\right)+\mathcal{H}_4\right],
\end{multline}
where,
\begin{multline}
\mathcal{H}_1=3 r_0 \left(r_0^2-4 A^2\right) \left(4 A^2 \left(6 A^2 r^2+8 A^4-r^4\right)\right.\\\left.
+\left(r^3-4 A^2 r\right)^2 \log \left(\frac{r^2}{r^2-4 A^2}\right)\right),
\end{multline}
\begin{equation}
    \mathcal{H}_2=\left(r \log \left(\frac{r^2}{r^2-4 A^2}\right)-r_0 \log \left(\frac{r_0^2}{r_0^2-4 A^2}\right)\right),
\end{equation}
\begin{equation}
\mathcal{H}_3=24 \pi ^2 \alpha  A^2 r^3 r_0^2\left(2 r_0^2+r^2-8A^2\right),
\end{equation}
%\begin{equation}
%\mathcal{H}_4=\pi ^2 \beta  r r_0 \left(r^2-4 A^2\right)^2 %\left(4 A^2-r_0^2\right) \left(r_0^3-4 r^3\right),
%\end{equation}
\begin{equation}
\mathcal{H}_4=3 r r_0^2 \left(r^2-4 A^2\right)^2 \left(4 A^2-r_0^2\right) \log \left(\frac{r_0^2}{r_0^2-4 A^2}\right).
\end{equation}
The graphical behavior of EoS parameters for this case has been shown in fig. \ref{fig:61}.\\
The expressions of NEC for this particular case can be obtained as
\begin{multline}\label{4cc2}
NEC1:\frac{1}{\mathcal{H}_5}\left[-128 A^8 \left(r \left(12 \pi ^2 \alpha r_0^2-3\right)+6 r_0\right)\right.\\\left.
+64 A^6 \left(6 \pi ^2 \alpha  r r_0^2 \left(2 r^2+r_0^2\right)-3 \left(r^3+r r_0^2-r_0^3\right)\right) \right.\\\left.
+24 A^4 r^3 \left(-4 \pi ^2 \alpha  r_0^2 \left(r^2+2 r_0^2\right)+r^2+4 r_0^2\right)
+4 A^2 r^5 r_0^2 \right.\\\left.
\times \left(6\pi ^2 \alpha r_0^2-3\right)-\mathcal{H}_4\right].
\end{multline}
\begin{multline}\label{4cc3}
NEC2: \frac{1}{2\mathcal{H}_5}\left[128 A^8 \left(2 \pi ^2 r_0 \left(6 \alpha  r_0\right)-3\right)+192 A^6 \right.\\\left.
\times \left(-2 \pi ^2 \alpha  r_0^2 \left(2 r^2+r_0^2\right)+r^2-3 r r_0+r_0^2\right)
+24 A^4 r \right.\\\left.
\times \left(4 \pi ^2 \alpha  r_0^2\left(r^3+2 r r_0^2\right)+4 r^2 r_0-r^3-4 r r_0^2+6 r_0^3\right)\right.\\\left.
+4 A^2 r^3 r_0^2 \left(r \left(3-6\pi^2 \alpha r_0^2\right)-6 r_0\right)+\mathcal{H}_6 \right],
\end{multline}
where
\begin{equation}
\mathcal{H}_5=24 \pi ^2 A^2 r^4 \left(r^2-4 A^2\right)^2 \left(4 A^2 r_0-r_0^3\right),
\end{equation}
\begin{multline}
\mathcal{H}_6=3 r_0 \left(r^2-4 A^2\right)^2 \left(4 A^2-r_0^2\right) \\
\times \left(r_0 \log \left(\frac{r_0^2}{r_0^2-4 A^2}\right)-2 r \log \left(\frac{r^2}{r^2-4 A^2}\right)\right).
\end{multline}
At $r=r_0$ (wormhole's throat), the above radial NEC reduces to
\begin{multline}
NEC1\mid_{r=r_0}=-\frac{1}{\mathcal{H}_7}\left[32 A^6+8 \pi ^2 \alpha  r_0^2 \left(A r_0^2-4 A^3\right)^2 \right.\\\left.
+24 A^4 r_0^2-4 A^2 r_0^4+\left(r_0^3-4 A^2 r_0\right)^2 \log \left(\frac{r_0^2}{r_0^2-4 A^2}\right)\right]
\end{multline}
where, $\mathcal{H}_7=8 \pi ^2 A^2 r_0^4 \left(r_0^2-4 A^2\right)^2.$ From the above expression, it is obvious that the NEC for radial pressure is violated at the wormhole's throat. However, for tangential pressure, NEC is satisfied. The graphical representations of NEC are depicted in Fig. \ref{fig:5=62}. One may also notice that for an increase of the model parameter $\alpha$, the contributions for the violation of NEC become higher. Further, we have studied the behavior of DEC in Fig. \ref{fig:63}. We observed that radial DEC is satisfied, whereas tangential DEC is violated in the entire space-time. SEC's situation was the same as in the previous two cases.

\subsubsection{$A$ variable and $D$ constant}
For this particular case, we replace $A$ by wormhole radius $r$ and considered the sphere radius $A$ as constant.
Comparing the Eqs. \eqref{14} and \eqref{4c1} and solving  the differential equation one can get the shape function as follows:
\begin{multline}\label{4c3}
    b(r)=-\frac{1}{8 \pi ^2 \alpha  D^4}\left[\frac{r}{6} \left(D^4 \left(\frac{3}{D^2-4 r^2}+8 \pi ^2 \beta  r^2\right)\right.\right.\\\left.\left.
    +27 D^2+4 r^2\right)+D^2 r \log \left(\frac{D^2}{D^2-4 r^2}\right)\right.\\\left.
    +\frac{5}{4} D^3 (\log (2 r-D)-\log (D+2 r))\right]+c_3.
\end{multline}
to find the integrating constant $c_3$, we are applying throat condition $b(r)=r_0$ at $r=r_0$:
\begin{multline}\label{4c4}
c_3=r_0+\frac{1}{8 \pi ^2 \alpha  D^4}\left[\frac{r_0}{6} \left(D^4 \left(\frac{3}{D^2-4 r_0^2}+8 \pi ^2 \beta  r_0^2\right)\right.\right.\\\left.\left.
    +27 D^2+4 r_0^2\right)+D^2 r_0 \log \left(\frac{D^2}{D^2-4 r_0^2}\right)\right.\\\left.
    +\frac{5}{4} D^3 (\log (2 r_0-D)-\log (D+2 r_0))\right].
\end{multline}
Inserting Eq. \eqref{4c3} into the Eq. \eqref{4c4}, we get the shape function given by
\begin{multline}\label{4c5}
b(r)=\frac{1}{96} \left[\frac{2 r}{\alpha }\left(-\frac{4 r^2}{\pi ^2 D^4}-\frac{3}{\pi ^2 \left(D^2-4 r^2\right)}-\frac{27}{\pi ^2 D^2}\right.\right.\\\left.\left.
-8 \beta  r^2\right)-\frac{3}{\pi ^2 \alpha  D^2} \left(\mathcal{K}_1+\mathcal{K}_4\right)+\frac{8 r_0^3}{\alpha }\left(2 \beta +\frac{1}{\pi ^2 D^4}\right)\right.\\\left.
+6 r_0 \left(\frac{1}{\pi ^2 \alpha }\left(\frac{1}{D^2-4 r_0^2}+\frac{9}{D^2}\right)+16\right)\right],
\end{multline}
where,
\begin{multline}
 \mathcal{K}_1=5D(\log (2 r-D)-\log (D+2 r)\\
 -\log (2 r_0-D)+\log (D+2 r_0)).
\end{multline}
Here also, we realized that the obtained $b(r)$ disrespects the asymptotically flatness condition. It happens because of the free parameter $\beta$. For $\beta\rightarrow 0$, the flatness condition $\frac{b(r)}{r}\nrightarrow 0$ will satisfy. Therefore the shape function can be obtained for vanishing $\beta$ from Eq. \eqref{4c5}
\begin{multline}\label{4cc5}
b(r)=\frac{1}{96} \left[\frac{2 r}{\alpha }\left(-\frac{4 r^2}{\pi ^2 D^4}-\frac{3}{\pi ^2 \left(D^2-4 r^2\right)}-\frac{27}{\pi ^2 D^2}\right)\right.\\\left.
-\frac{3}{\pi ^2 \alpha  D^2} \left(\mathcal{K}_1+\mathcal{K}_4\right)+\frac{8 r_0^3}{\alpha \pi ^2 D^4}+6 r_0 \right.\\\left.
\times \left(\frac{1}{\pi ^2 \alpha }\left(\frac{1}{D^2-4 r_0^2}+\frac{9}{D^2}\right)+16\right)\right].
\end{multline}

\begin{figure}[h]
\includegraphics[scale=0.6]{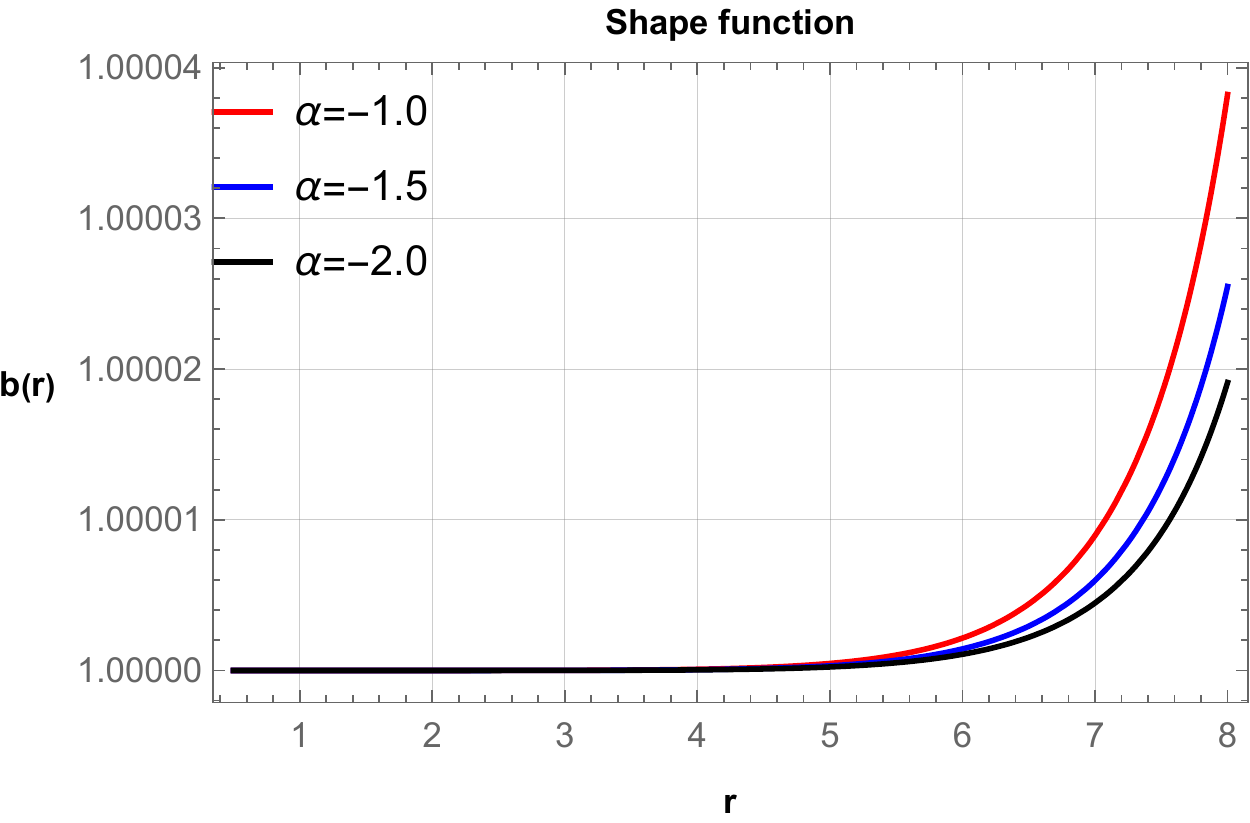}
\includegraphics[scale=0.6]{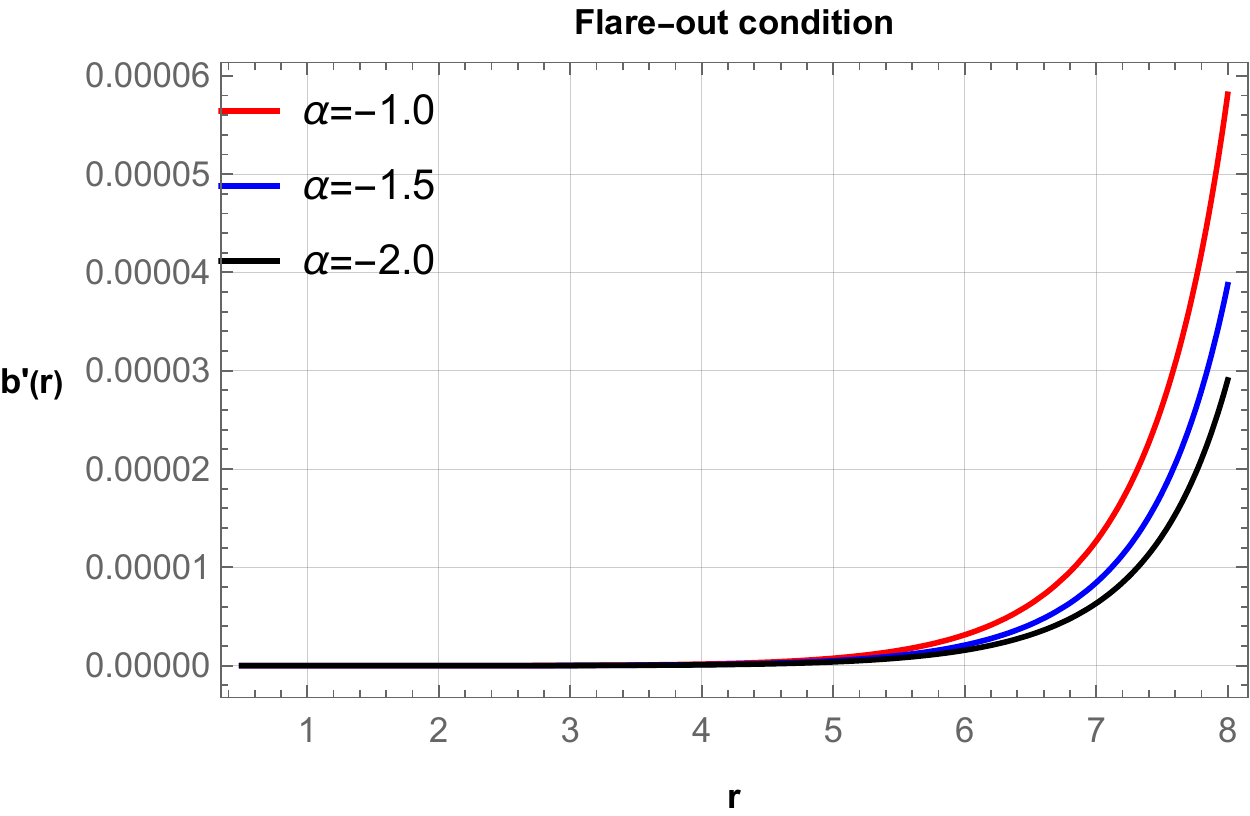}
\caption{Shape function and flare-out condition for two sphere (case-2) with with $r_0=1$ and $D=20$.}
\label{fig6aaa}
\end{figure}
\begin{figure}[h]
    \includegraphics[scale=0.6]{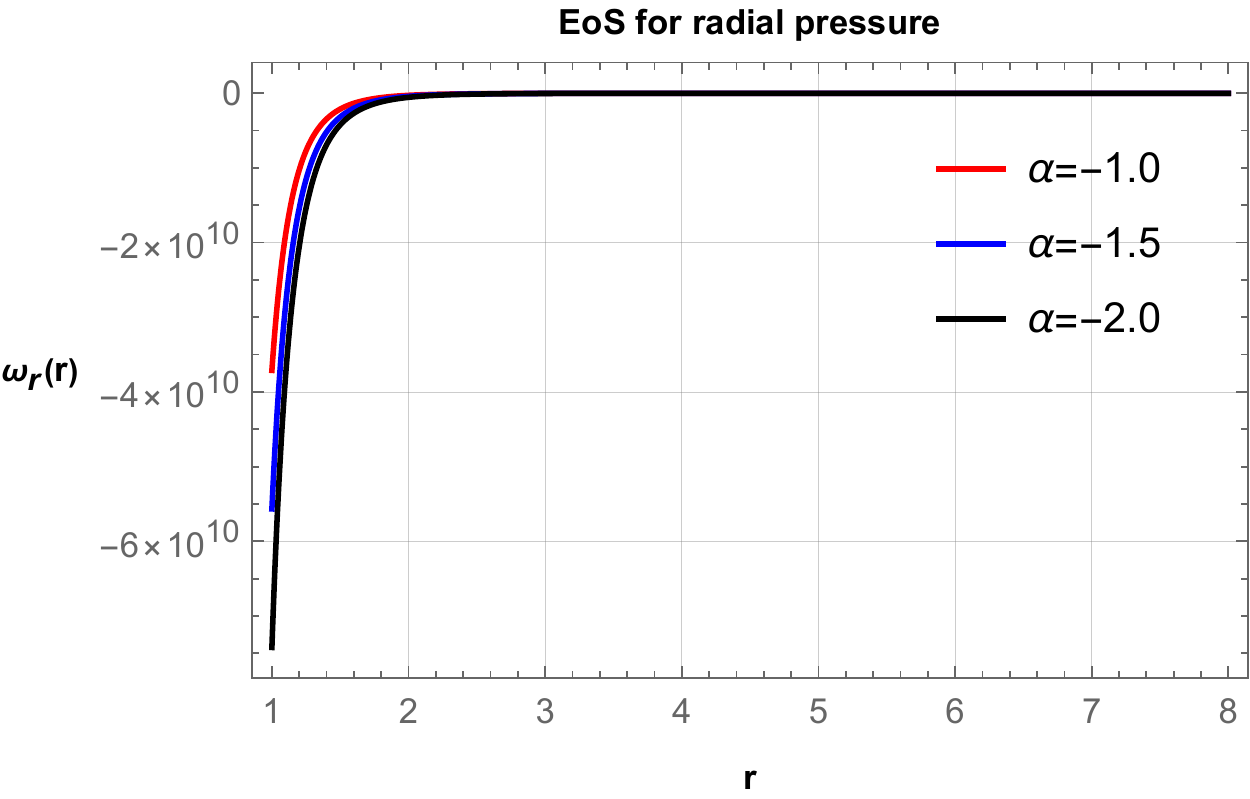}
    \includegraphics[scale=0.6]{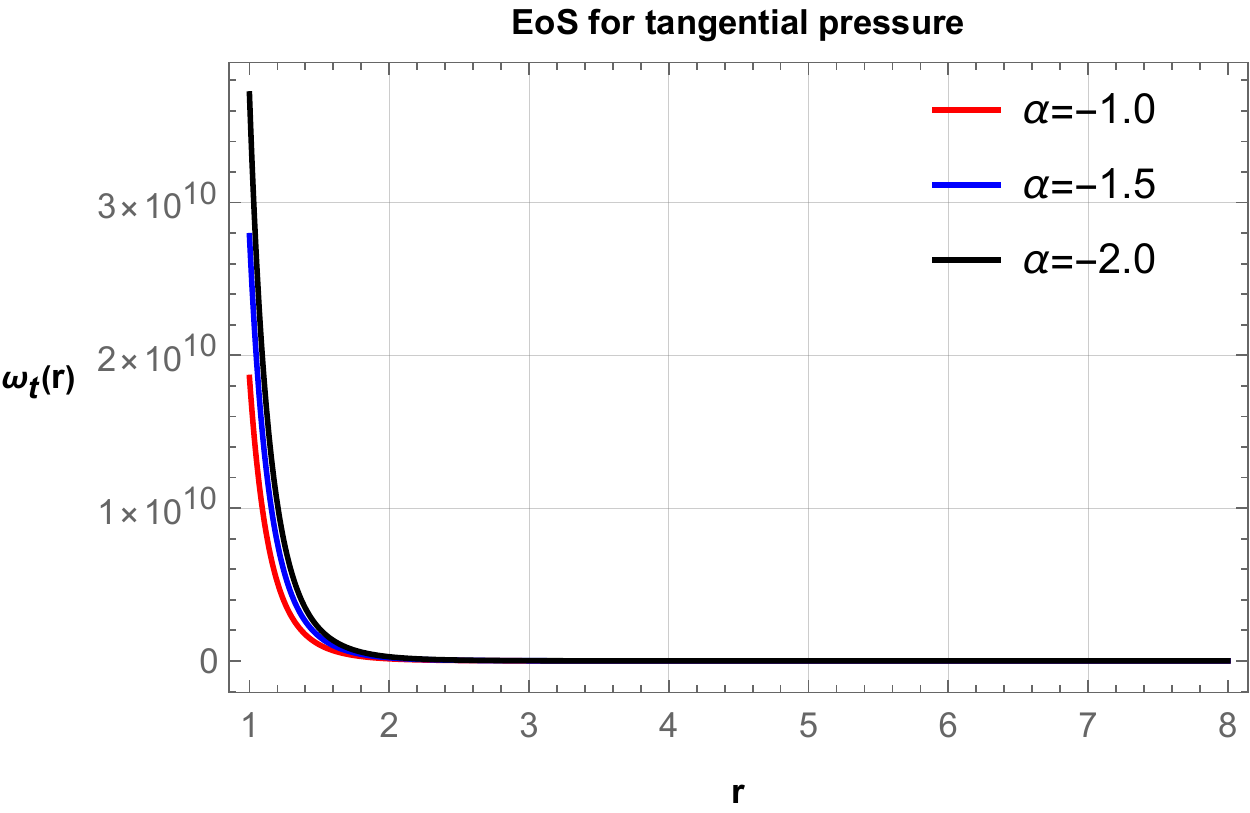}
    \caption{ The EoS parameter $\omega$ for two sphere (case-2) with $r_0=1$, $D=20$ and vanishing $\beta=0$.}
    \label{fig:51}
\end{figure}

In Fig. \ref{fig6aaa}, we have plotted the shape function $b(r)$ and flaring out condition $b^{'}(r_0)<1$. We identify that these behavior satisfy for negative $\alpha$ only.\\
The radial and tangential pressure components for this case can be read as
\begin{multline}\label{4c6}
P_r=\frac{1}{96 r^3}\left[\frac{8 \left(r^3-r_0^3\right)}{\pi ^2 D^4}+\frac{6}{\pi ^2}\left(\frac{r}{D^2-4 r^2}-\frac{r_0}{D^2-4 r_0^2}\right)\right.\\\left.
+\frac{3}{\pi ^2 D^2}\left(\mathcal{K}_4+\mathcal{K}_1\right)+\frac{54 (r-r_0)}{\pi ^2 D^2}-96\alpha r_0\right],
\end{multline}
\begin{multline}\label{4c7}
P_t=\frac{1}{192} \left[\mathcal{K}_2+\frac{8 \left(2 r^3+r_0^3\right)}{\pi ^2 D^4 r^3}+\frac{1}{\pi ^2 D^2 r^3}\right.\\\left.
\times \left(12 r_0 \log \left(\frac{D^2}{D^2-4 r_0^2}\right)
-3\mathcal{K}_1\right)+\mathcal{K}_3\right],
\end{multline}
where,
\begin{multline}
\mathcal{K}_2=\frac{6}{\pi ^2}\left(\frac{\frac{r_0}{D^2-4 r_0^2}-\frac{10 r}{D^2-4 r^2}}{r^3}+\frac{8}{\left(D^2-4 r^2\right)^2}\right),
\end{multline}
\begin{multline}
\mathcal{K}_3=\frac{6}{\pi ^2 D^2}\left(\frac{16}{D^2-4 r^2}+\frac{9 r_0}{r^3}\right)+\frac{96 \alpha  r_0}{r^3},
\end{multline}
\begin{multline}
\mathcal{K}_4=4 r \log \left(\frac{D^2}{D^2-4 r^2}\right)-4 r_0 \log \left(\frac{D^2}{D^2-4 r_0^2}\right).
\end{multline}
Considering Eqs. \eqref{4c1}, \eqref{4c6} and \eqref{4c7}, we have illustrated the plot for radial and tangential EoS parameters in Fig. \ref{fig:51}.\\
The expressions for NEC for two sphere (case-2) can be written as
%\begin{widetext}
\begin{multline}\label{4c8}
NEC1:\quad \frac{1}{96} \left[-\mathcal{K}_2-\frac{8 \left(2 r^3+r_0^3\right)}{\pi ^2 D^4 r^3}+\frac{1}{\pi ^2 D^2 r^3}\right.\\\left.
\times\left(3\mathcal{K}_1-12 r_0 \log \left(\frac{D^2}{D^2-4 r_0^2}\right)\right)-\mathcal{K}_3\right],
\end{multline}
\begin{multline}\label{4c9}
NEC2:\quad \frac{1}{192} \left[\frac{2}{r^3}\left(48 \alpha  r_0+\frac{24 r \left(D^2-5 r^2\right)}{\pi ^2 \left(D^2-4 r^2\right)^2}\right.\right.\\\left.\left.
+\frac{3 r_0}{\pi ^2 \left(D^2-4 r_0^2\right)}
\right)+\frac{8 \left(r_0^3-4 r^3\right)}{\pi ^2 D^4 r^3}
+\frac{3}{\pi ^2 D^2 r^3}\right.\\\left.
\times \left(-4 r \log \left(\frac{D^2}{D^2-4 r^2}\right)+5 D \log (2 r-D)\right.\right.\\\left.\left.
-(\mathcal{K}_1+\mathcal{K}_4)\right)
+\frac{6}{\pi ^2 D^2}\left(\frac{9 (r_0-2 r)}{r^3}-\frac{16}{D^2-4 r^2}\right)\right.\\\left.
-\frac{15 \log (2 r-D)}{\pi ^2 D r^3}\right].
\end{multline}
%\end{widetext}
At throat, the above expressions for NEC reduces to
\begin{multline}\label{4c10}
NEC1\mid_{r=r_0}=-\frac{1}{\mathcal{K}_5}\left[4 \left(2 \alpha \pi ^2 D^8-16 \alpha \pi ^2 D^6 r_0^2 \right.\right.\\\left.\left.
+D^4 r_0^2 \left(32 \alpha \pi^2 r_0^2-1\right)+6 D^2 r_0^4+8 r_0^6\right)+\mathcal{K}_6\right],
\end{multline}
\begin{multline}\label{4c11}
 NEC2\mid_{r=r_0}=\frac{1}{2\mathcal{K}_5}\left[4 \left(2 \pi ^2 \alpha  D^8-16 \pi ^2 \alpha  D^6 r_0^2\right.\right.\\\left.\left.
 +32 \pi ^2 \alpha  D^4 r_0^4+D^4 r_0^2
 -6 D^2 r_0^4-8 r_0^6\right)-\mathcal{K}_6\right],
\end{multline}
where,\\
$\mathcal{K}_5=8 \pi ^2 D^4 r_0^2 \left(D^2-4 r_0^2\right)^2,$\\
$\mathcal{K}_6=\left(D^3-4 D r_0^2\right)^2 \log \left(\frac{D^2}{D^2-4 r_0^2}\right).$\\
From Eqs. \eqref{4c10} and \eqref{4c11} it is evident that for negative values of $\alpha$, NEC is violated for tangential pressure at the throat whereas satisfied for radial pressure.\\
\begin{figure}[h]
    \includegraphics[scale=0.6]{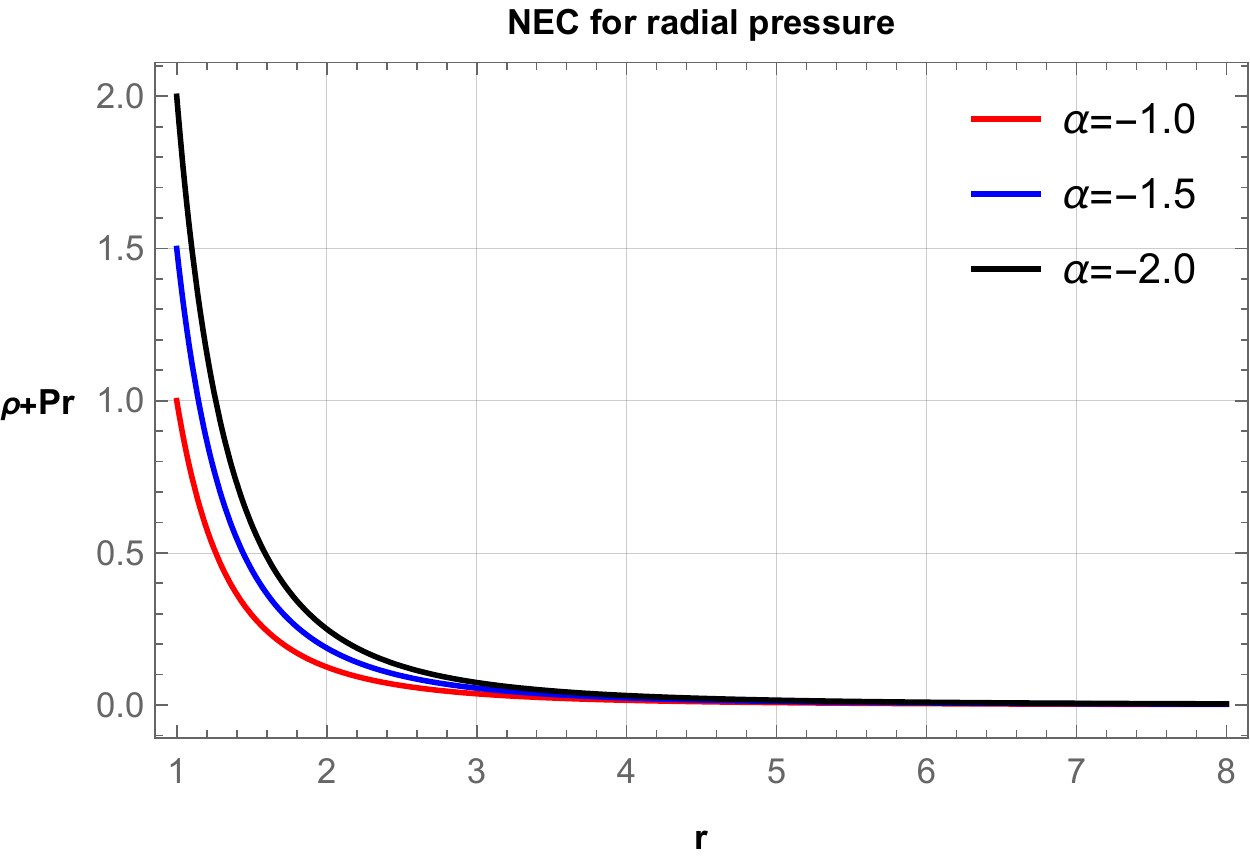}
    \includegraphics[scale=0.6]{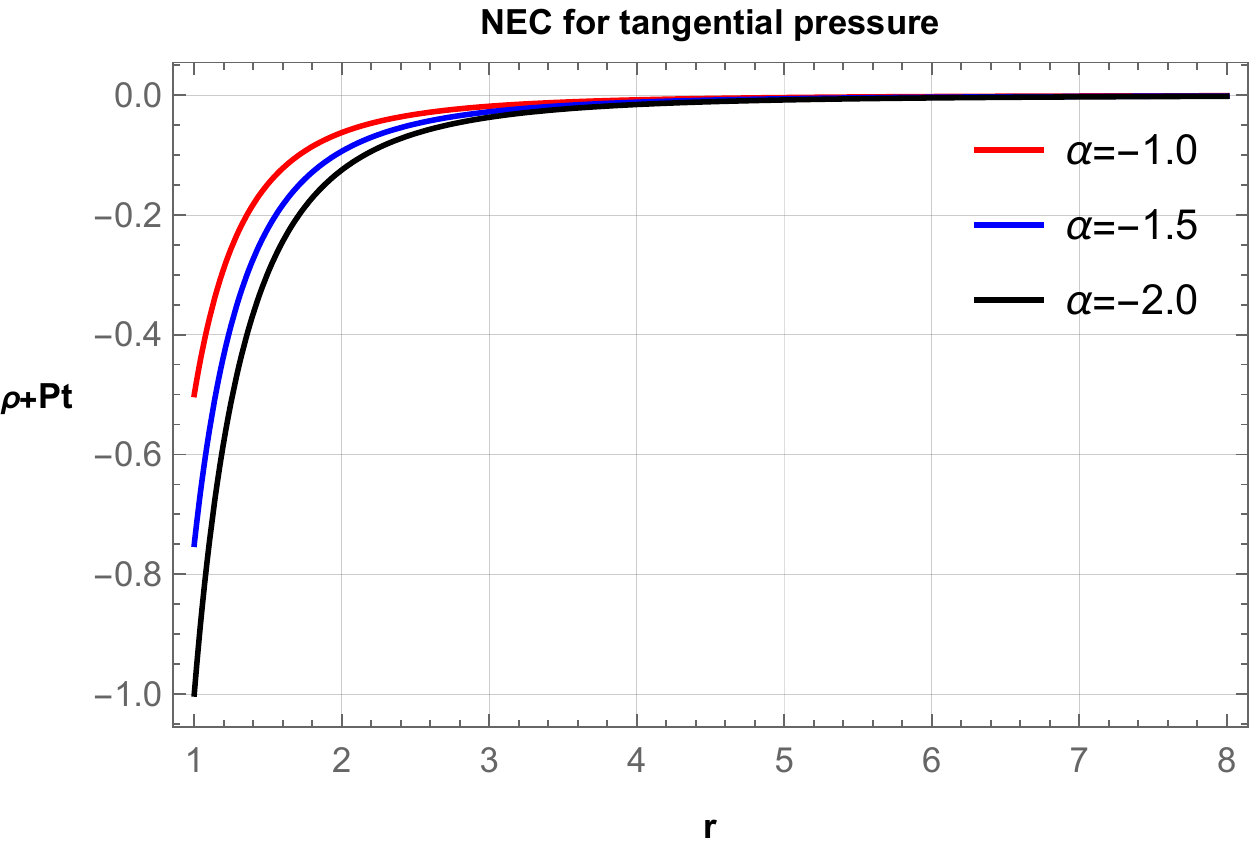}
    \caption{The variation of NEC against $r$ for two sphere (case-2) with $r_0=1$ and $D=20$.}
    \label{fig:52}
\end{figure}
\begin{figure}[h]
    \includegraphics[scale=0.6]{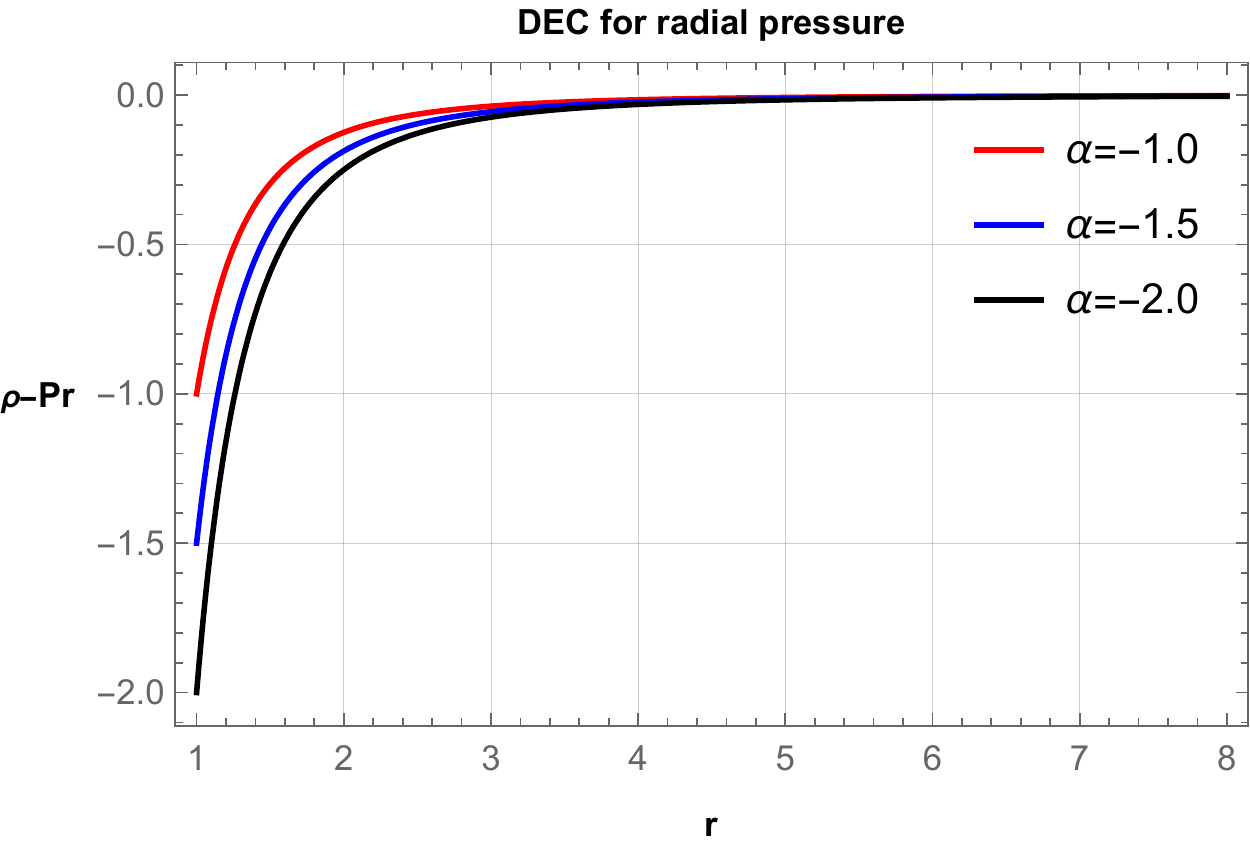}
    \includegraphics[scale=0.6]{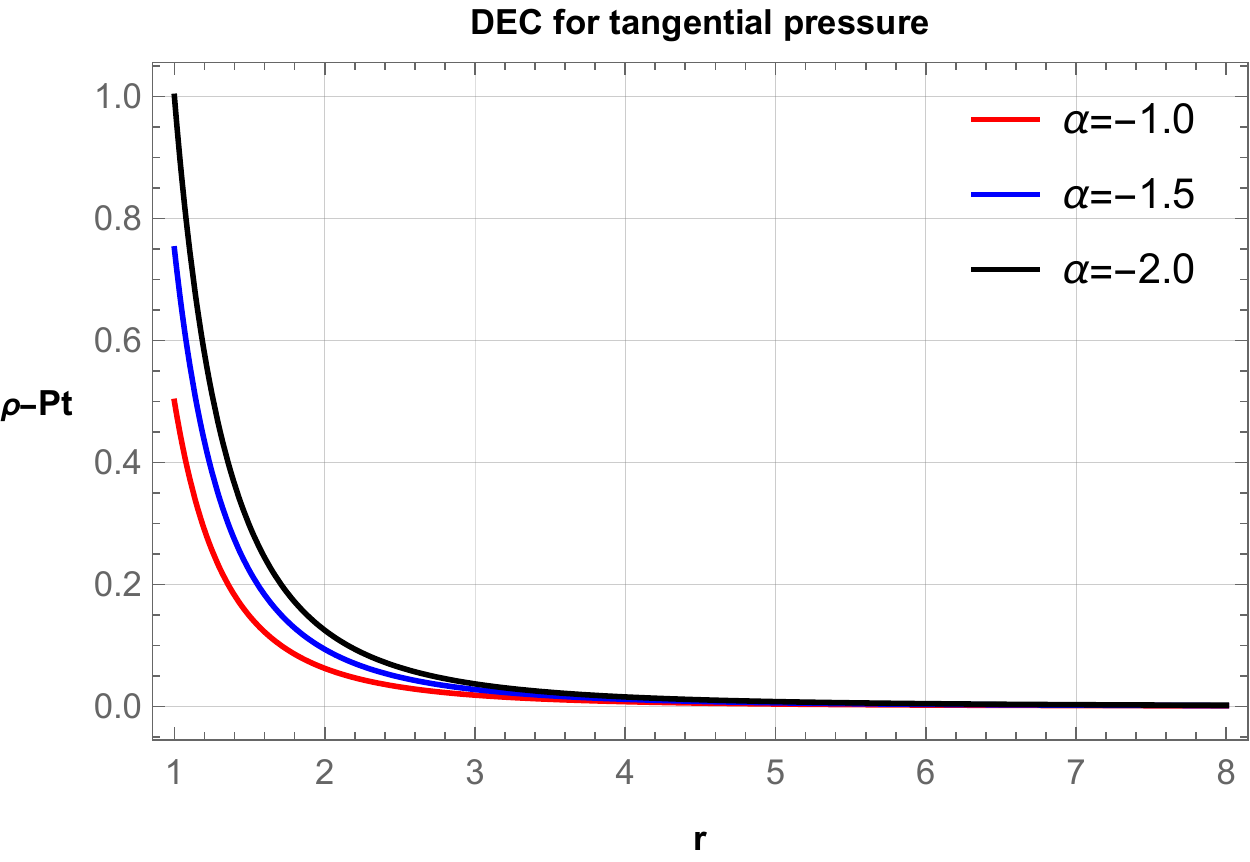}
    \caption{The variation of DEC against $r$ for two sphere (case-2) with $r_0=1$ and $D=20$.}
    \label{fig:53}
\end{figure}
Figs. \ref{fig:52} and \ref{fig:53} shows the characteristics of NEC and DEC for both pressures. It was mentioned that NEC is disrespected for tangential pressure. But, DEC is satisfied for tangential pressure whereas violated for radial pressure. SEC situation is the same as the previous two cases.
\section{Equilibrium conditions}\label{sec4}
In this particular section, we consider generalized Tolman-Oppenheimer-Volkoff (TOV) equation to investigate the stability of our obtained wormhole solutions. The generalized TOV equation can be define as \cite{Islam,Kuhfittig/2020}
\begin{equation}\label{5a}
-\frac{dP_{r}}{dr}-\frac{\phi^{'}(r)}{2}(\rho+P_{r})+\frac{2}{r}(P_{t}-P_{r})=0,
\end{equation}
Naturally, the generalized TOV equation \eqref{5a} gives the information of the equilibrium condition for the wormhole
subject to the hydrostatic $(F_H)$ and gravitational $(F_G)$ force with another force called anisotropic $(F_A)$ force (because of anisotropic matter).\\
Hence, the equation \eqref{5a} takes the following form
\begin{equation}\label{5a2}
    F_H+F_G+F_A=0,
\end{equation}
where,
\begin{equation}\label{5a3}
F_H=-\frac{dP_{r}}{dr},    
\end{equation}
\begin{equation}\label{5a4}
F_G=-\frac{\phi^{'}}{2}(\rho+P_{r}), 
\end{equation}
\begin{equation}\label{5a5}
 F_A=\frac{2}{r}(P_{t}-P_{r}).   
\end{equation}
Since in our study we considered the constant redshift function (tidal less force). Hence, the gravitational force will not be appear here, i.e., $F_G=0$.\\
Thus, the Eq. \eqref{5a2} reduces to
\begin{equation}\label{5a6}
F_A+F_H=0.
\end{equation}
\begin{figure}[h]
    \centering
    \includegraphics[scale=0.6]{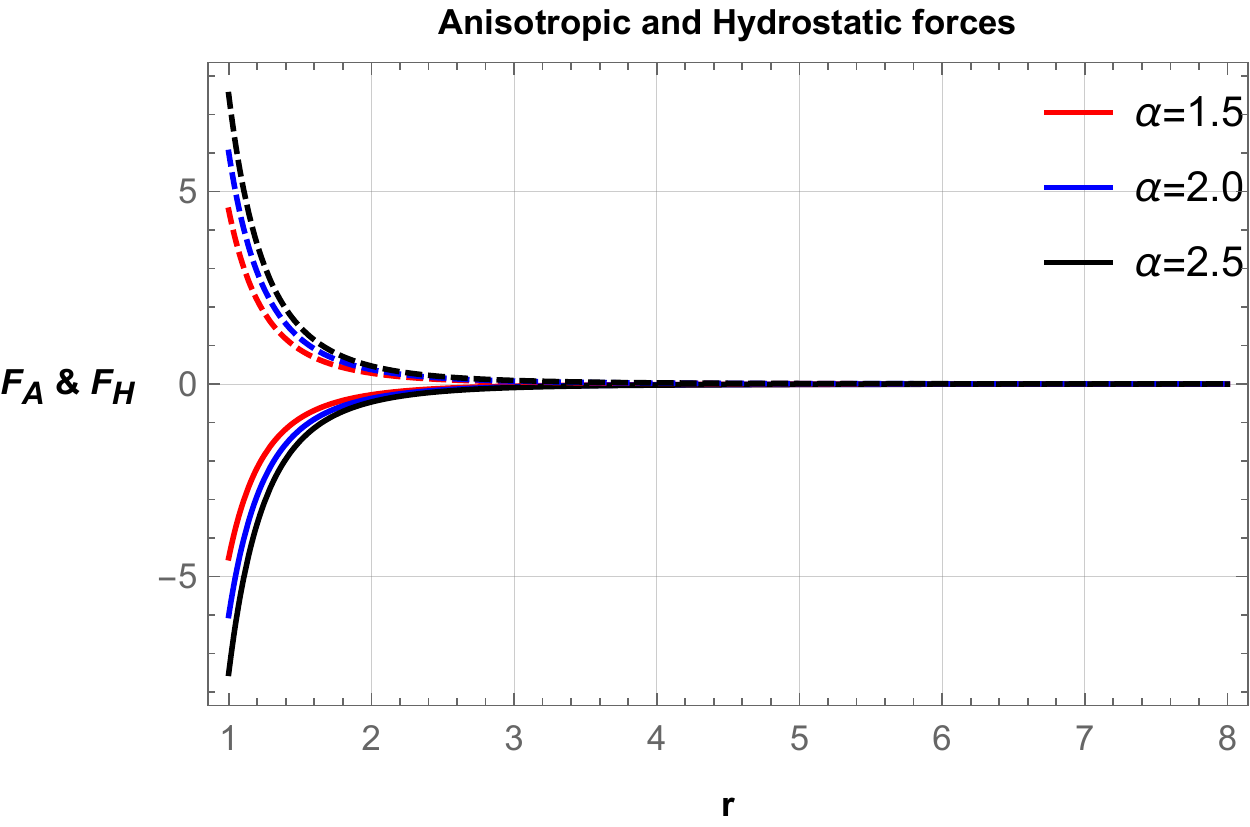}
    \caption{Hydrostatic (plane lines) and anisotropic (dashed lines) forces for parallel plates with $r_0=1$.}
    \label{fig:8a}
\end{figure}
\begin{figure}[h]
    \centering
    \includegraphics[scale=0.6]{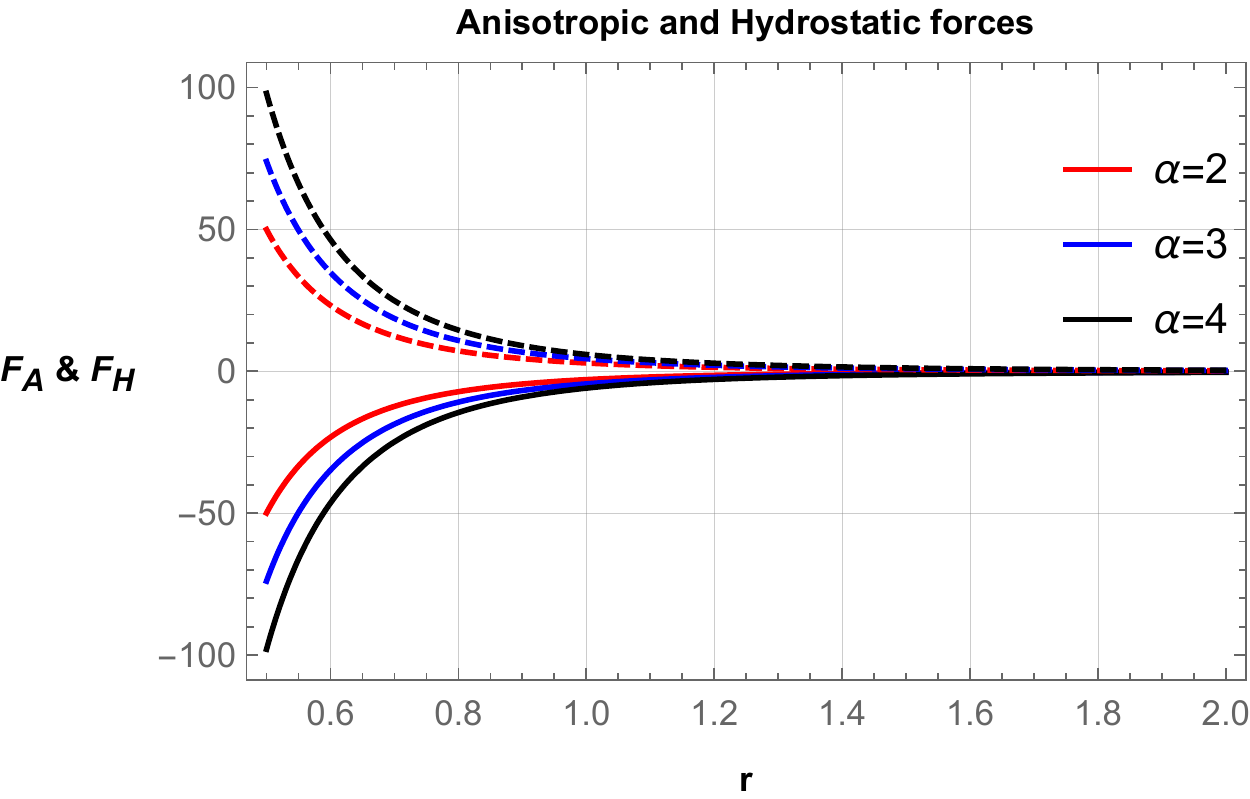}
    \caption{Hydrostatic (plane lines) and anisotropic (dashed lines) forces for parallel cylinder with $r_0=0.5$.}
    \label{fig:8aa}
\end{figure}
\begin{figure}[h]
    \centering
    \includegraphics[scale=0.6]{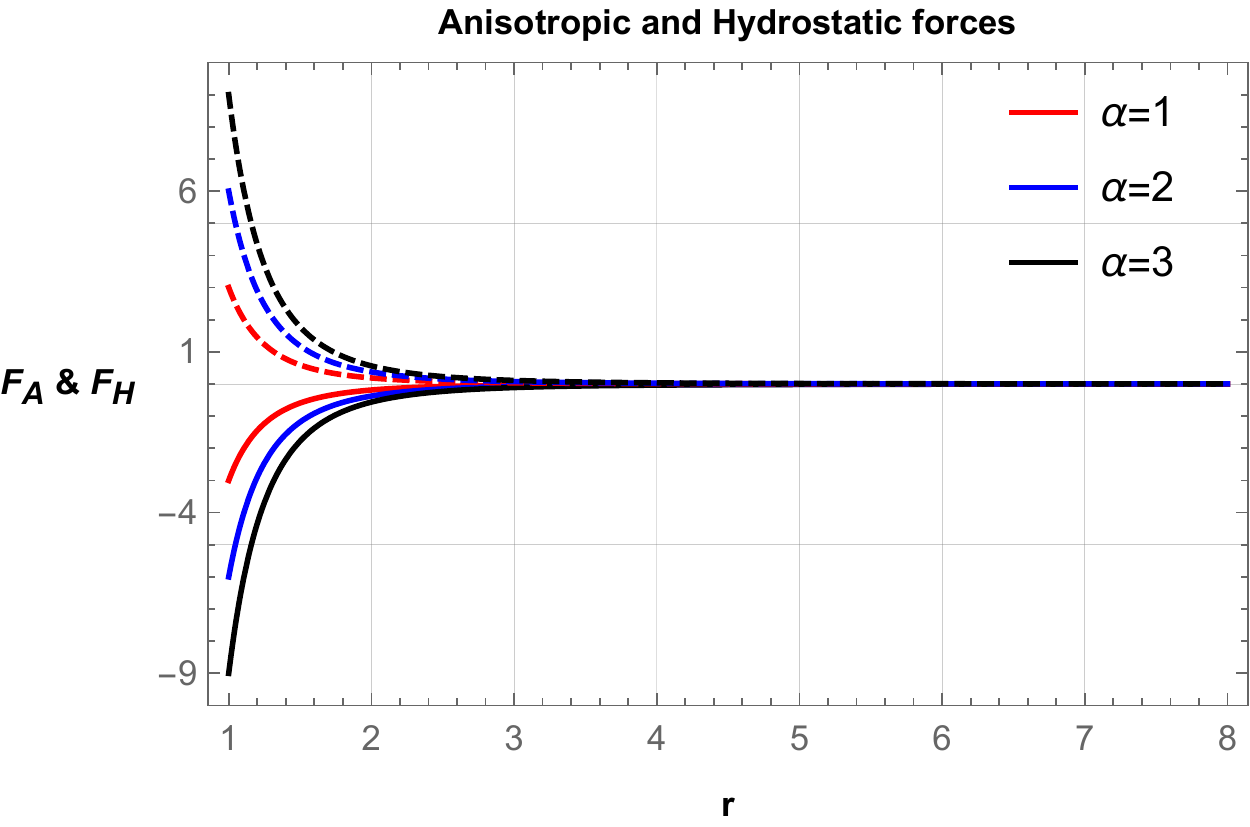}
    \caption{Hydrostatic (plane lines) and anisotropic (dashed lines) forces for two sphere (case-1) with $r_0=1$ and $A=0.05$.}
    \label{fig:8aaaa}
\end{figure}
\begin{figure}[h]
    \centering
    \includegraphics[scale=0.6]{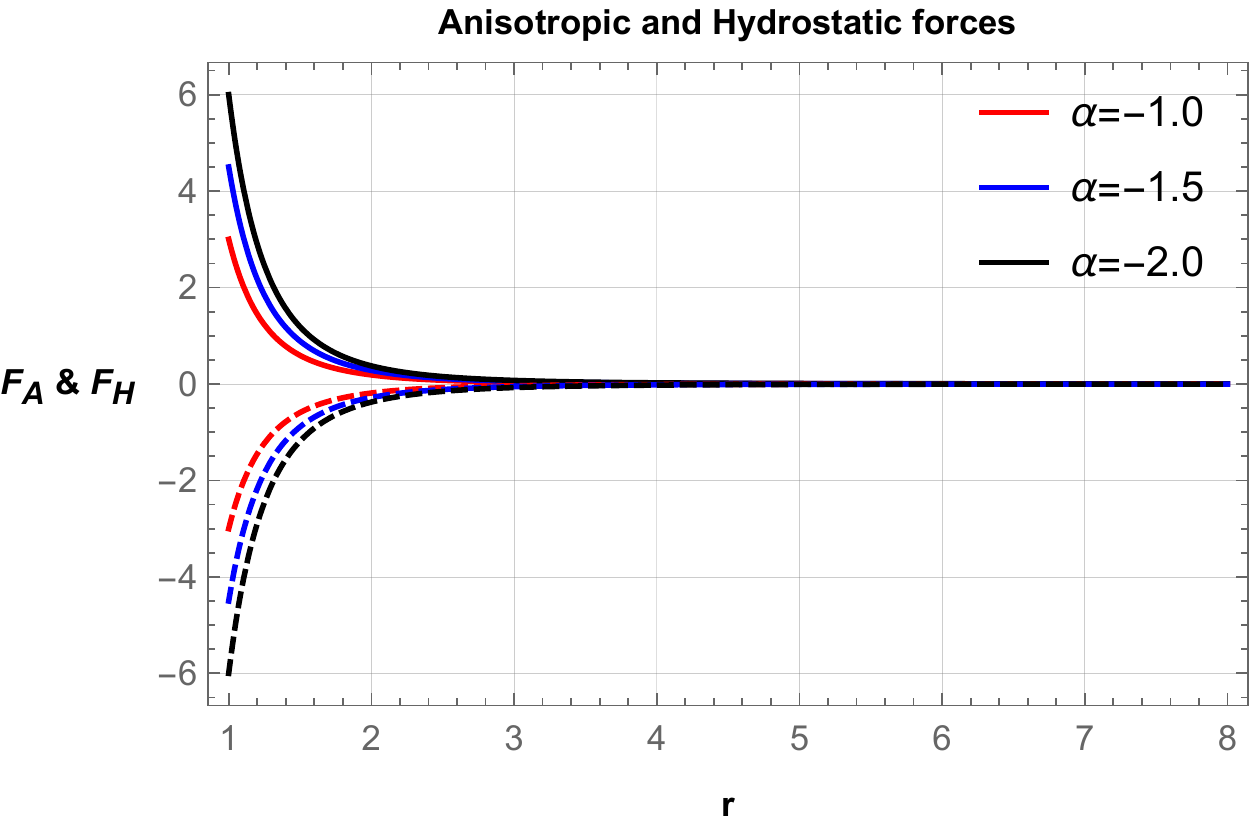}
    \caption{Hydrostatic (plane lines) and anisotropic (dashed lines) forces for two sphere (case-2) with $r_0=1$ and $D=20$.}
    \label{fig:8aaa}
\end{figure}
%Considering Eqs. (\ref{4a2}, \ref{4a6} and \ref{4a7}) (for parallel plates), Eqs. (\ref{4b4}-\ref{4b6}) (for parallel cylinder) and Eqs. (\ref{4c1}, \ref{4c6} and \ref{4c7}) (two spheres cases) into account,
Using the appropriate expressions, we have plotted the graph for hydrostatic ($F_H$) and anisotropic ($F_A$) forces for each case in Figs. \ref{fig:8a}-\ref{fig:8aaa}. It is observed that anisotropic $F_A$ force shows positive behavior, whereas hydrostatic $F_H$ force shows negative behavior for the first two cases (parallel plates and parallel cylinder). For the two spheres cases, we noticed that both forces behave in opposite behavior. One may find that these forces are identical but opposite to each other, resulting in the equilibrium of the solutions. Thus, it is safe to conclude that our obtained wormhole solutions for each case are stable. There are few Refs. \cite{Karar,Ilyas,Mustafa/2022} where the authors studied deeply on this interesting topic.\\

\section{Conclusions}\label{sec5}
In this study, we have investigated the Casimir effect on wormhole space-time in modified symmetric teleparallel gravity. We have taken three different systems which we can think mimic the actual picture of the wormhole. Firstly, we have considered the parallel plate as done by Garattini \cite{Garattini} and extended it into the Morris-Thorne wormhole in GR. Jusufi et al. \cite{Channuie} extended Garattini's work and studied the effect of GUP in the Casimir wormhole by considering three GUP relations such as Kempf, Mangano, and Mann (KMM) model, the Detournay, Gabriel, and Spindel (DGS) model, and type II model. Further, Tripathy \cite{Tripathy} investigated the Casimir wormholes with the above KMM and DGS models in modified $f(R,T)$ gravity. However, in our study, besides parallel plate, We have also taken two different systems called parallel cylinder and two spheres to elaborate on the fact that wormhole stability indeed works in different scenarios.\\
%Apart from considering Casimir wormhole in $f(Q)$ gravity we also considered two other systems that is parallel cylinder and two spheres where the Casimir force can be approximately calculated and used that expression to calculate the energy density($\rho$). As far our knowledge the Casimir wormhole beyond parallel plate has not been studied in literature. The advantage of studying Casimir wormhole in parallel cylinder and two sphere has two main advantage, first of all these are more practically achievable and can indeed be helpful two design in some experiment about wormhole done in near future, and secondly as we can see changing the system keeps the wormhole stable but the value changes of $b(r)$ we can use the information we got from the graph to extrapolate what should be the optimal radius of cylinder or sphere should be and what should be the optimal distance between two plates in case for practical consideration. \\
From the Casimir energy densities of the three systems, we have integrated the energy density of modified gravity to obtain the shape function of the wormhole metric. Then, we studied the obtained shape function for each system graphically. It is observed that the shape functions we got satisfy the flare-out conditions under asymptotic background. Also, we noticed the effect of modified gravity on shape functions. For the first case, one can visualize from the graph that an increase in the value of the modified gravity parameter outcomes in an increase in the shape function with a radial distance from the wormhole throat. Also, for the parallel cylinder case, we observed the increase in the value of the parameter results in an increase in the shape function. For the last case, it was noticed that a decrease in the value of the modified gravity parameter decreases the shape function. However, the impact of the modified gravity is insignificant at a radial distance nearer to the throat. One crucial observation is that the shape functions obtained from the Casimir sources do not follow the asymptotically flatness condition. We noticed that it happens due to the presence of free parameter $\beta$. Thus we are bound to vanish the parameter $\beta$ from the obtained shape functions in order to satisfy the flatness condition. Further, we have studied the EoS parameters for radial and tangential pressure to the Casimir energy density for each case. It can be observed that for the first two cases, the radial EoS parameter shows an increasing function against radial distance, whereas the tangential EoS parameter is a decreasing function. For the two spheres cases, for $case-1$, both EoS parameters indicate the same behaviors as the previous two cases, whereas for $case-2$, both parameters show opposite behaviors. It came to notice that the modified gravity parameter impacts the behavior of both EoS parameters to a large extent, at least at distances from the wormhole throat. Moreover, one can observe that the graphs of EoS parameters, $\omega_r$ and $\omega_t$ are diverging when $r\rightarrow \infty$. The reason being that from Eq. \eqref{4a2}, one can see that the Casimir force decays as $\frac{1}{r^4}$, thus at $r\rightarrow \infty$, $\rho$ goes to zero. However, we can see from Eqs. \eqref{4a6} and \eqref{4a7}, the pressure does not go to zero when $r\rightarrow \infty$ (for parallel plates case). As a consequence, $\omega_i=\frac{P_i}{\rho}$ (where $i$ is either $r$ or $t$) are unbounded as $r\rightarrow \infty$ (as $\rho$ tends to zero). However, one can avoid such a consequence by noting that our Casimir plates are placed very near to the wormhole throat and are for engineering purposes to supply. So we can not take the values of $\rho$ and $P_i$ for arbitrary $r$. For that reason, we have calculated all the quantities and shown their behaviors in the vicinity of the throat because at $r\rightarrow \infty$, the components of the stress-energy tensor will no longer be valid.\\
Furthermore, we have investigated the NEC, DEC, and SEC at the wormhole's throat with a radius of $r_0$ for each case. It was observed that for the first two cases, NEC was violated for any $\alpha>0$ (one may refer to the Eqs. \eqref{4a14} and \eqref{4b9}). For the third case, radial NEC was disrespected for positive values of $\alpha$ $(case-1)$, and for $case-2$, NEC was violated for tangential pressure for any $\alpha<0$. DEC was violated for tangential pressure only for the first two cases and $(case-1)$ of two spheres case, whereas for the $case-2$, it was violated for radial pressure. For the SEC, we noticed that it depends on the constraint $\beta$ only. For $\beta=0$, SEC will vanish. This situation aligns with the work of Ref. \cite{Hassan/2021}. Furthermore, we used an interesting tool called Tolman-Oppenheimer-Volkoff (TOV) equation to verify whether the obtained wormhole solutions are stable or unstable. Throughout our observations, we found that our obtained wormhole solutions are stable.\\
Recently, one interesting work was done by Sokoliuk et al. in \cite{Sokoliuk}, where they studied the Casimir wormholes in the framework of $f(R)$ gravity. Also, the authors of Ref. \cite{Channuie1} investigated charged wormholes supported by Casimir energy with and without GUP corrections. The above authors used Casimir density for parallel plates to study the wormholes. However, in this work, we look at two more Casimir energy densities besides parallel plate density. Thus, this study may provide new insight into the field of modified theories of gravity. Also, it would be interesting to investigate Casimir wormholes with GUP corrected relations (KMM, DGS, and type-II) in modified symmetric teleparallel gravity in the near future.

\section*{Data Availability Statement}
There are no new data associated with this article.
%%%%%
\section*{Acknowledgments}

ZH acknowledges Department of Science and Technology (DST), Government of India, New Delhi, for awarding a Senior Research Fellowship (File No. DST/INSPIRE Fellowship/2019/IF190911). SG acknowledges Council of Scientific and Industrial Research (CSIR), Government of India, New Delhi, for junior research fellowship (File no.09/1026(13105)/2022-EMR-I). PKS acknowledges National Board for Higher Mathematics (NBHM) under Department of Atomic Energy (DAE), Govt. of India for financial support to carry out the Research project No.: 02011/3/2022 NBHM(R.P.)/R\&D II/2152 Dt.14.02.2022. The work of KB was partially supported by the JSPS KAKENHI Grant Number 21K03547. We are very much grateful to the honorable referee and to the editor for the illuminating suggestions that have significantly improved our work in terms
of research quality, and presentation. 
%%%%%


\begin{thebibliography}{52}
\footnotesize
\bibitem{Padmanabhan} T. Padmanabhan, Gravitation, first ed., \textit{Cambridge University Press}, Cambridge, (2010).
\bibitem{Misner/1973} C. W. Misner, K. S. Thorne and J. A. Wheeler, Gravitation, first ed., W. H. Freeman, San Francisco, (1973).
\bibitem{Dyson} F. W. Dyson, A. S. Eddington, C. Davidson, \textit{Phil. Trans. Roy. Soc. Lond. A} \textbf{220}, 291-333 (1920).
\bibitem{Suzuki} A. Ishihara, Y. Suzuki, T. Ono, T. Kitamura, H. Asada, \textit{Phys. Rev. D} \textbf{94}, 084015 (2016).
\bibitem{Abbott} B. P. Abbott et al. [LIGO Scientific and Virgo], \textit{Phys. Rev. Lett.} \textbf{116}, 061102 (2016).
\bibitem{Schwarzschild} K. Schwarzschild, Sitzungsber. Preuss. Akad. Wiss. Berlin (Math. Phys.) \textbf{1916}, 189-196 (1916).
\bibitem{Akiyama/2019} K. Akiyama et al. [Event Horizon Telescope], \textit{Astrophys. J. Lett.} \textbf{875}, L1 (2019).
\bibitem{Akiyama/2021} K. Akiyama, et al. \textit{Astrophys. J. Lett.} \textbf{910}, L13 (2021).
\bibitem{Flamm} L. Flamm, \textit{Physikalische Zeitschrift} \textbf{17}, 48 (1916).
\bibitem{Rosen} A. Einstein, N. Rosen, \textit{Phys. Rev.} \textbf{48}, 73 (1935).
\bibitem{Visser} M. Visser, Lorentzian Wormholes: From Einstein to Hawking, first ed., American Institute of Physics, New York,
(1996).
\bibitem{Thorne/1988} M. S. Morris, K. S. Thorne, \textit{Am. J. Phys.} \textbf{56}, 395-412 (1988).
\bibitem{Hochberg} D. Hochberg, M. Visser, \textit{Phys. Rev. D} \textbf{58}, 044021 (1998).
\bibitem{Yurtsever} M.S. Morris, K.S. Thorne, U. Yurtsever, \textit{Phys. Rev. Lett.} \textbf{61}, 1446 (1988).
\bibitem{Kim} S. W. Kim, \textit{J. Korean Phys. Soc.} \textbf{63}, 1887-1891 (2013).
\bibitem{Simeone} E. F. Eiroa and C. Simeone, \textit{Phys. Rev. D} \textbf{81}, 084022 (2010).
\bibitem{Gao} P. Gao, D. L. Jafferis, A. C. Wall, \textit{J. High Energy Phys.} \textbf{12}, 151 (2017).
\bibitem{Milekhin} J. Maldacena, A. Milekhin, \textit{Phys. Rev. D} \textbf{103},  066007 (2021).
\bibitem{Perlmutter} S. Perlmutter et al. \textit{Astrophys. J.} \textbf{517}, 565 (1999).
\bibitem{Komatsu}  E. Komatsu et al. \textit{Astrophys. J. Suppl. Ser.} \textbf{192}, 18 (2011).
\bibitem{Riess} A. G. Riess, et al. \textit{Astrophys. J.} \textbf{659}, 98 (2007).
\bibitem{Suzuki/2012} N. Suzuki, et al. \textit{Astrophys. J.} \textbf{746}, 85 (2012).
\bibitem{Buchdahl/1970} H.A. Buchdahl, \textit{Mon. Not. R. Astron. Soc.} \textbf{150}, 1-8 (1970).
\bibitem{Starobinsky/1980} A. Starobinsky, \textit{Phys. Lett. B} \textbf{91}, 99-102 (1980).
\bibitem{Lobo/2011} T. Harko, F. S. N. Lobo, S. Nojiri, S. D. Odintsov, \textit{Phys. Rev. D} \textbf{84}, 024020 (2011).
\bibitem{Lobo/2009} F. S. N. Lobo, M. A. Oliveira, \textit{Phys. Rev. D} \textbf{80}, 104012 (2009).
\bibitem{Bronnikov/2010} K. A. Bronnikov, M. V. Skvortsova, A. A. Starobinsky, \textit{Gravit. Cosmol.} \textbf{16}, 216 (2010).
\bibitem{Kar/2003} M. Visser, S. Kar, N. Dadhich, \textit{Phys. Rev. Lett.} \textbf{90}, 201102 (2003).
\bibitem{Banerjee} K. Jusufi, N. Sarkar, F. Rahaman, A. Banerjee, S. Hansraj, \textit{Eur. Phys. J. C} \textbf{78}, 349 (2018).
\bibitem{Bambi/2015} N. Tsukamoto, C. Bambi, \textit{Phys. Rev. D} \textbf{91}, 084013 (2015).
\bibitem{Ovgun} A. Ovgun, K. Jusufi, I. Sakalli, \textit{Phys. Rev. D} \textbf{99}, 024042 (2019).
\bibitem{Sahoo1} P. K. Sahoo, P. H. R. S. Moraes, P. Sahoo, \textit{Eur. Phys. J. C} \textbf{78}, 46 (2018).
\bibitem{Karar1} F. Rahaman, I. Karar, S. Karmakar, S. Ray, \textit{Phys. Lett. B} \textbf{746}, 73 (2015).
\bibitem{Shaikh} R. Shaikh, \textit{Phys. Rev. D} \textbf{98}, 024044 (2018).
\bibitem{Nedkova} G. Gyulchev, P. Nedkova, V. Tinchev, S. Yazadjiev, \textit{Eur. Phys. J. C} \textbf{78}, 544 (2018).
\bibitem{Tsukamoto} N. Tsukamoto, T. Harada, K. Yajima, \textit{Phys. Rev. D} \textbf{86}, 104062 (2012).
\bibitem{Amir} K. Jusufi, A. Banerjee, G. Gyulchev, M. Amir, \textit{Eur. Phys. J.C} \textbf{79}, 28 (2019).
\bibitem{Bahamonde2} S. Bahamonde, M. Jamil, P. Pavlovic, M. Sossich, \textit{Phys. Rev. D} \textbf{94}, 044041 (2016).
\bibitem{Bahamonde} S. Bahamonde, U. Camci, S. Capozziello, M. Jamil, \textit{Phys. Rev. D} \textbf{94}, 084042 (2016).
\bibitem{casimir}  H. B. G. Casimir,  D. Polder, \textit{Phys. Rev.} \textbf{73}, 360-372 (1948).
\bibitem{lifshitz} I. E. Dzyaloshinskii, E. M. Lifshitz, Lev P Pitaevskii,  \textit{Soviet Physics Uspekhi.}, \textbf{4}, 153 (1961).
\bibitem{experiment} S. K. Lamoreaux,  \textit{ Phys. Rev. Lett.}, \textbf{78}, 5-8 (1997).
\bibitem{bressi} G. Bressi, G. Carugno, R. Onofrio, G. Ruoso \textit{Phys. Rev. Lett.}, \textbf{88}, 041804 (2002).
\bibitem{PFA} B. V. Derjaguin, I. I. Abrikosova, Sov. Phys. JETP
\textbf{3}, 819 (1957); B. V. Derjaguin, Sci. Am. \textbf{203}, 47 (1960).
\bibitem{Jimenez}  J. B. Jimenez, L. Heisenberg, T. Koivisto, \textit{Phys. Rev. D}, \textbf{98}, 044048 (2018).
\bibitem{Mandal/2020} S. Mandal, D. Wang,  P. K. Sahoo, \textit{Phys. Rev. D} \textbf{102}, 124029 (2020).
\bibitem{Mandal/2020a} S. Mandal, P.K. Sahoo., J.R.L. Santos
\textit{Phys. Rev. D} \textbf{102}, 024057 (2020).
\bibitem{Zhao} D. Zhao, \textit{Eur. Phys. J. C} \textbf{82}, 303 (2022).
\bibitem{Frusciante} N. Frusciante \textit{Phys. Rev. D} \textbf{103}, 0444021 (2021).
\bibitem{Harko/2018a} T. Harko et al. \textit{Phys. Rev. D} \textbf{98}, 084043 (2018).
\bibitem{Khyllep1} W. Khyllep, A. Paliathanasis, J. Dutta 
\textit{Phys. Rev. D} \textbf{103}, 103521 (2021).
\bibitem{Anagnostopoulos} F. K. Anagnostopoulos, S. Basilakos, E. N. Saridakis, \textit{Phys. Lett. B} \textbf{822}, 136634 (2021).
\bibitem{Fell} F. D'Ambrosio, S. D. B. Fell, L. Heisenberg, S. Kuhn, \textit{Phys. Rev. D} \textbf{105}, 024042 (2022).
\bibitem{Hassan1} Z. Hassan, S. Mandal, P.K. Sahoo, \textit{Fortschritte Phys.} \textbf{69}, 2100023 (2021).
\bibitem{Mustafa} G. Mustafa, Z. Hassan, P.H.R.S. Moraes, P.K. Sahoo, \textit{Phys. Lett. B} \textbf{821}, 136612 (2021).
\bibitem{Sharma1} U. K. Sharma, Shweta, A. K. Mishra, \textit{Int. J. Geom. Methods Mod. Phys.} \textbf{19}, 2250019 (2021).
\bibitem{Banerjee1} A. Banerjee, A. Pradhan, T. Tangphati, F. Rahaman, \textit{Eur. Phys. J. C} \textbf{81}, 1031 (2021).
\bibitem{Mustafa2} S. Mandal, G. Mustafa, Z. Hassan, P. K. Sahoo, \textit{Phys. Dark Univ.} \textbf{35}, 100934 (2022).
\bibitem{Lazkoz} R. Lazkoz, et al. \textit{Phys. Rev. D} \textbf{100}, 104027 (2019).
\bibitem{Lin} Rui-Hui Lin and Xiang-Hua Zhai, \textit{Phys. Rev. D} \textbf{103}, 124001 (2021).
\bibitem{Solanki1} R. Solanki, S. K. J. Pacif, A. Parida, P. K. Sahoo, \textit{Phys. Dark Univ.} \textbf{32}, 100820 (2021).
\bibitem{Wang2} W. Wang, H. Chen, T. Katsuragawa, \textit{Phys. Rev. D} \textbf{105},  024060 (2022).
\bibitem{Koivisto} J. B. Jimenez, L. Heisenberg, T. Koivisto, and S. Pekar, \textit{Phys. Rev. D} \textbf{101}, 103507 (2020).
\bibitem{2a} M.S. Morris, K.S. Throne, U. Yurtseve, \textit{Phys. Rev. Lett.} \textbf{61}, 1446 (1988).
\bibitem{Solanki} R. Solanki, A. De, P. K. Sahoo, \textit{Phys. Dark Univ.} \textbf{36}, 101053 (2022).
\bibitem{paddy} T. Padmanabhan, Quantum Field Theory The Why, What and How, first ed, Springer, (2016).
\bibitem{zee} A. Zee, Quantum Field Theory in a Nutshell, second ed, Princeton University Press, (2010).
\bibitem{Santos} A. C. L. Santos, C. R. Muniz, L. T. Oliveira, \textit{Int. J. Mod. Phys. D} \textbf{30}, 2150032 (2021).
\bibitem{Garattini} R. Garattini, \textit{Eur. Phys. J. C} \textbf{79}, 951 (2019).
\bibitem{Hassan/2021} Z. Hassan, G. Mustafa, P. K. Sahoo, \textit{symmetry}, \textbf{13}, 1260 (2021).
\bibitem{cylinder} F. C. Lombardo, F. D. Mazzitelli, P. I. Villar \textit{J. Phys. A: Math. Theor.} \textbf{41}, 164009 (2008).
\textit{J. Phys. A: Math. Theor.}, \textbf{41}, 164009 (2008).
\bibitem{mazzitelli}F. D. Mazzitelli, M. J. Sanchez, N. N. Scoccola, J. V. Stecher \textit{Phys. Rev. A} \textbf{67}, 013807(2003).
\bibitem{Sphere} J. P. Straley, E. B. Kolomeisky \textit{ J. Phys.: Condens. Matter}, \textbf{29}, 143002 (2017).
\bibitem{kardar}T. Emig, N. Graham, R. L. Jaffe, M. Kardar \textit{Phys. Rev. Lett.}, \textbf{99}, 170403 (2007).
\bibitem{klich}O. Kenneth, I. Klich \textit{Phys. Rev. B}, \textbf{78}, 014103 (2008).
\bibitem{Islam} F. Rahaman, P. K. F. Kuhfittig and N. Islam, \textit{Eur. Phys. J. C} \textbf{74}, 2750 (2014).
\bibitem{Kuhfittig/2020}  P. K. F. Kuhfittig, \textit{Fund. J. Mod. Phys.} \textbf{14}, 23 (2020).
\bibitem{Karar} F. Rahaman, S. Karmakar, I. Karar,  S. Ray, \textit{Phys. Lett. B} \textbf{746}, 73 (2015).
\bibitem{Ilyas} Z. Yousaf, M. Ilyas,  M. Zaeem-ul-Haq Bhatti, \textit{Eur. Phys. J. Plus} \textbf{132}, 268 (2017).
\bibitem{Mustafa/2022} G. Mustafa, Z. Hassan, P.K. Sahoo, \textit{Ann. Phys.} \textbf{437}, 168751 (2022).
\bibitem{Channuie} K. Jusufi, P. Channuie, M. Jamil, \textit{Eur. Phys. J. C} \textbf{80}, 127 (2020).
\bibitem{Tripathy} S. K. Tripathy, \textit{Phys. Dark Univ.} \textbf{31}, 100757 (2021).
\bibitem{Sokoliuk} O. Sokoliuk, A. Baransky, P. K. Sahoo, \textit{Nuclear Phys. B} \textbf{980}, 115845 (2022). 
\bibitem{Channuie1} D. Samart, T. Tangphati, P. Channuie, \textit{Nuclear Phys. B} \textbf{980}, 115848 (2022).

\end{thebibliography}
\end{document}